%% file: main.tex
\documentclass[journal]{IEEEtran}

\input{preamble}

\begin{document}

\title{Machine Learning in Compiler Optimisation}
\author{
    Zheng Wang and Michael O'Boyle

    \thanks{Z. Wang is with MetaLab, School of Computing and Communications, Lancaster University, U. K. E-mail: z.wang@lancaster.ac.uk}%
    \thanks{M. O'Boyle is with School of Informatics, University of Edinburgh, U. K. E-mail: mob@inf.ed.ac.uk}
}

\maketitle

\input{abstract}

\input{intro}

\input{overview}
\input{motivation}
\input{methodology}
\input{models}
\input{feature}
\input{scope}
\input{challenges}
\input{conclusions}

\IEEEpeerreviewmaketitle

\bibliographystyle{IEEEtran}
\bibliography{zheng,mob}
\balance

%

\end{document}

%% file: preamble.tex
\usepackage{csquotes}
\usepackage{xcolor}
\usepackage{color}

\usepackage[normalem]{ulem}
\usepackage{booktabs}
\usepackage{tabularx}
\usepackage{hhline}
\usepackage{xspace}
\usepackage{graphicx}
\usepackage{subfig}
\usepackage{tabularx,colortbl}

\usepackage{multirow}
\usepackage{balance}

\usepackage{listings}
 \lstdefinelanguage[OpenCL]{C}[ANSI]{C}
{morekeywords={__kernel,kernel,__local,local,__global,global,%
__constant,constant,__private,private,%
char2,char3,char4,char8,char16,%
uchar2,uchar3,uchar4,uchar8,uchar16,%
short2,short3,short4,short8,short16,%
ushort2,ushort3,ushort4,ushort8,ushort16,%
int2,int3,int4,int8,int16,%
uint2,uint3,uint4,uint8,uint16,%
long2,long3,long4,long8,long16,%
ulong2,ulong3,ulong4,ulong8,ulong16,%
float2,float3,float4,float8,float16,%
image2d_t,image3d_t,sampler_t,event_t,size_t,%
bool2,bool3,bool4,bool8,bool16,%
half2,half3,half4,half8,half16,%
quad,quad2,quad3,quad4,quad8,quad16,%
complex,imaginary},%
}%

\newcommand\etal{\emph{et al.}\xspace}
\newcommand\GP{\texttt{GP}\xspace}

\newcommand\ANN{\texttt{ANN}\xspace}
\newcommand\NUMA{\texttt{NUMA}\xspace}
\newcommand\SVM{\texttt{SVM}\xspace}
\newcommand\SVMs{\texttt{SVMs}\xspace}
\newcommand\NN{\texttt{NN}\xspace}
\newcommand\KNN{\texttt{KNN}\xspace}
\newcommand\RL{\texttt{RL}\xspace}
\newcommand\PCA{\texttt{PCA}\xspace}

\newcommand\DNNs{\texttt{DNNs}\xspace}

\newcommand\FFT{\texttt{FFT}\xspace}

\definecolor{Gray}{gray}{0.9}

%% file: abstract.tex
\begin{abstract}
In the last decade, machine learning based compilation has moved from an an obscure research niche to a  mainstream activity. In this
article, we describe the relationship between machine learning and compiler optimisation and introduce the main concepts of features,
models, training and deployment. We then provide a comprehensive survey  and provide a road map for the wide variety of different
research areas. We conclude with a discussion on open issues in the area and potential research directions. This paper provides both an
accessible introduction to the fast moving area of machine learning based compilation and a detailed bibliography of its main
achievements.
\end{abstract}


%% file: intro.tex
\section{Introduction}

``Why would anyone want to use machine learning to build a compiler?'' It's a view expressed by many colleagues over the last decade.
Compilers translate programming languages written by humans into binary executable by computer hardware. It is a serious subject studied
since the 50s \cite{chipps1956mathematical,sheridan1959arithmetic,mcilroy1960macro} where correctness is critical and caution is a by-word.
Machine-learning on the other hand is an area of artificial intelligence aimed at detecting and predicting patterns. It is a dynamic field
looking at subjects as diverse as galaxy classification~\cite{gauci2010machine} to predicting elections based on tweeter
feeds~\cite{schoen2013power}. When an open-source machine learning compiler was announced by IBM in 2009 \cite{milepostgcc:slashdot}, some
wry slashdot commentators picked up on the AI aspect, predicting the start of sentient computers, global net and the war with machines
from the Terminator film series.

In fact as we will see in this article, compilers and machine learning are a natural fit and have developed into an established  research
domain. 

\subsection{It's all about optimization}

Compiler have two jobs -- translation and optimisation. They must first translate programs into binary correctly. Secondly they have to
find the most efficient translation possible. There are many different correct translations whose performance varies significantly. The
vast majority of research and engineering practices is focussed on this second goal of performance, traditionally misnamed optimisation.
The goal was misnamed because in most cases, till recently finding an optimal translation was dismissed as being too hard to find and an
unrealistic endeavour\footnote{In fact the term superoptimizer \cite{massalin1987superoptimizer} was coined to describe systems that tried
to find the optimum}. Instead it focussed on developing compiler heuristics to transform the code in the hope of improving  performance but
could in some instances damage it.

Machine learning predicts an outcome for a new data point based on prior data. In its simplest guise it can be considered a from of
interpolation.  This ability to predict based on prior information can be used to find the data point with the best outcome and is closely
tied to the area of optimisation. It is at this overlap of looking at code improvement as an optimisation problem and machine learning as a
predictor of the optima where we find machine-learning compilation.

Optimisation as an area, machine-learning based or otherwise, has been studied since the 1800s~\cite{ivory1825method,adcock1878problem}. An
interesting question is therefore why has has the convergence of these two areas taken so long? There are two fundamental reasons. Firstly,
despite the year-on year increasing potential performance of hardware, software is increasingly unable to realise it leading to a
software-gap. This gap has yawned right open with the advent of multi-cores (see also Section~\ref{sec:parallel_opt}). Compiler writers are
looking for new ways to bridge this gap.

Secondly, computer architecture evolves  so quickly, that it is difficult to keep up. Each generation has new quirks and compiler writers
are always trying to play catch-up. Machine learning has the desirable property of being automatic. Rather than relying on expert compiler
writers  to develop clever heuristics to optimise the code, we can let the machine learn how to optimise a compiler to make the machine run
faster, an approach sometimes referred to as
auto-tuning~\cite{datta2008stencil,ansel:cgo:2011,kurzak2016implementation,tsai2016performance}. Machine learning is, therefore,  ideally suited to making 
{\em any}  code optimization decision where the performance impact depends on the underlying platform. As described later in this paper,
it can be used for topics ranging from selecting the best compiler flags to determining how to map parallelism to processors.  

Machine learning is part of a tradition in computer science and compilation in increasing automation The 50s to 70s were spent trying to
automate compiler translation, e.g. lex for lexical analysis \cite{lesk1975lex} and yacc for parsing \cite{johnson1975yacc}, the last
decade by contrast has focussed on trying  to automating compiler optimisation. As we will see it is not ``magic" or a panacea for compiler
writers, rather it is another tool allowing automation of tedious aspects of compilation providing new opportunities for innovation. It
also brings compilation nearer to the standards of evidence based science. It introduces an experimental methodology where we separate out
evaluation from design and considers the robustness of solutions. Machine learning based schemes in general have the problem of relying on
black-boxes whose working we do not understand and hence trust. This problem is just as true for machine learning based compilers. In this
paper we aim to demystify machine learning based compilation and show it is a  trustworthy and exciting direction for compiler research.

\begin{figure*}
  \centering
  \includegraphics[width=\textwidth]{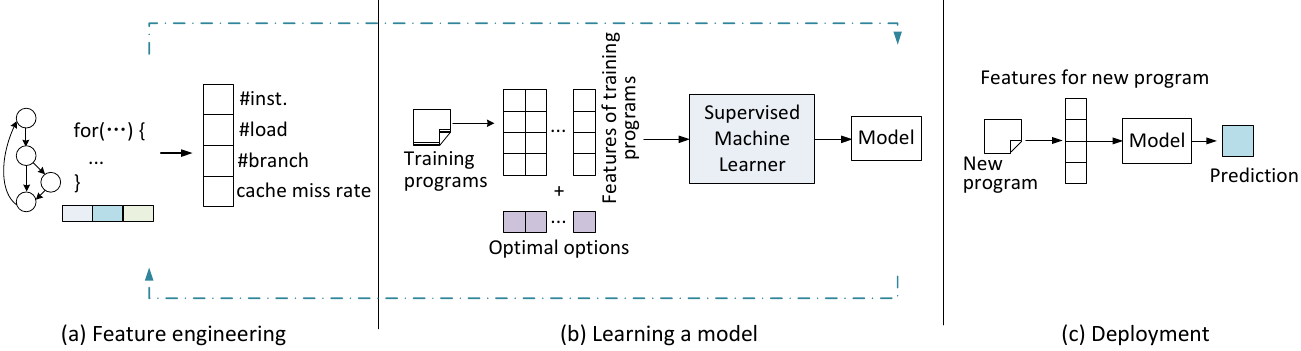}\\
  \caption{A generic view of supervised machine learning in compilers.
  }
  \label{fig:mlflow}
  \vspace{-3mm}
\end{figure*}

The remainder of this article is structured as follows. We first give an intuitive overview for machine learning in compilers in
Section~\ref {sec:overview}. We then describe how machine learning can be used to search for or to directly predict good compiler
optimizations in Section~\ref {sec:methodology}. This is followed by a comprehensive discussion in Section~\ref {sec:models} for a wide
range of machine learning models that have been employed in prior work. Next, in Section~\ref {sec:featureengineering}, we review how
previous work chooses quantifiable properties, or features, to represent programs. We discuss the challenges and limitations for applying
machine learning to compilation, as well as open research directions in Section~\ref{sec:discussion} before we summarise and conclude in
Section~\ref {sec:conclusion}.

%% file: overview.tex
\section{Overview of Machine Learning in Compilers \label{sec:overview}}
Given a program, compiler writers would like to know what compiler heuristic or optimisation to apply in order to make the code  better.
Better often means execute faster, but can also mean smaller code footprint or reduced power. Machine learning can be  used to build a
model used within  the compiler, that makes such decisions  for any given program.

There are two main stages involved: learning and deployment. The first stage learns the model based on training data, while the second uses
the model on new unseen programs. Within the learning stage, we  needs a way of representing programs in a systematic way. This
representation is known as the program features \cite{monsifrot2002machine}.

Figure~\ref{fig:mlflow} gives a intuitive view how machine learning can be applied to compilers. This process which includes feature
engineering, learning a model and deployment is described in the following sub-sections.

\subsection{Feature engineering}
Before we can learn anything useful about programs, we first need to be able to characterise them. Machine learning relies on  a set of
quantifiable  properties, or \emph{features}, to characterise the programs (Figure~\ref{fig:mlflow}a). There are many different features
that can be used. These include the static data structures extracted from the program source code or the compiler intermediate
representation (such as the number of instructions or branches), dynamic profiling information (such as performance counter values)
obtained through runtime profiling, or a combination of the both.

Standard machine learning algorithms typically work on fixed length inputs, so the selected properties will be summarised into a fixed
length \emph{feature vector}. Each element of the vector can be an integer, real or Boolean value. The process of feature selection and
tuning is referred as \emph{feature engineering}. This process may need to iteratively perform multiple times to find a set of high-quality
features to build a accurate machine learning model. In Section~\ref{sec:featureengineering}, we provide a comprehensive review of feature
engineering for the topic of program optimisation.

\subsection{Learning a model}
The second step is to use training data to derive  a model using a learning algorithm. This process is depicted in
Figure~\ref{fig:mlflow}b. Unlike other applications of machine learning, we typically generate our own training data using existing
applications or benchmarks. The compiler developer will select  training programs which are typical of the application domain. For each
training program, we calculate the feature values, compiling the program with different optimisation options, and running and timing the
compiled binaries to discover the best-performing option. This process produces, for each training program, a training instance that
consists of the feature values and the optimal compiler option for the program.

The compiler developer then feeds these examples to a machine learning algorithm to automatically build a model. The
learning algorithm’s job is to find from the training examples a correlation between the feature values and the optimal
optimisation decision. The learned model can then be used to predict, for a new set of features, what the optimal
optimisation option should be.

Because the performance of the learned model strongly depends on how well the features and training programs are chosen, so that
the processes of featuring engineering and training data generation often need to repeat multiple times.

\subsection{Deployment}
In the final step, the learned model is inserted into the compiler to predict the best optimisation decisions for new programs. This is
demonstrated in Figure~\ref{fig:mlflow}c. To make a prediction, the compiler first extracts the features of the input program, and then
feeds the extracted feature values to the learned model to make a prediction.

The advantage of the machine learning based approach is that the entire process of building the model can be easily
repeated whenever the compiler needs to target a new hardware architecture, operating system or application domain.
The model built is entirely derived from  experimental results and is hence evidence based.

%% file: motivation.tex
\subsection{Example \label{sec:motivation}}

\lstset{language=C}
\begin{figure}[t!]
	\centering %
	\subfloat[Original OpenCL kernel]{%
		\noindent\mbox{\parbox{\columnwidth}{%
\begin{lstlisting}[label=subfig:original]
kernel void square(global float* in, global float* out){
    int gid = get_global_id(0);
    out[gid] = in[gid] * in[gid];	
}
\end{lstlisting}%
		}}%
	}

    \subfloat[Code transformation with a coarsening factor of 2] {
        \noindent\mbox{\parbox{\columnwidth}{%
\begin{lstlisting}[label=subfig:transformed]
kernel void square(global float* in, global float* out){
    int gid = get_global_id(0);
    int tid0 = 2*gid + 0;
    int tid1 = 2*gid + 1;
    out[tid0] = in[tid0] * in[tid0];
    out[tid1] = in[tid1] * in[tid1];
}
\end{lstlisting}%
		}}%
    }

	\caption{An OpenCL thread coarsening example reproduced from \cite{Magni:2014:AOT:2628071.2628087}. The original OpenCL code is shown
at (a) where each thread takes the square of one element of the input array. When coarsened by a factor of two (b), each thread now
processes two elements of the input array. }
	\label{fig:coarsening}%
\end{figure}

As an example to illustrate these steps, consider thread coarsening~\cite{Unkule:2012:ARG:2259230.2259233} for GPU programs. This code
transformation technique works by giving multiple work-items (or work elements) to one single thread. It is similar to loop unrolling, but
applied across parallel work-items rather than across serial loop iterations.

Figure~\ref{fig:coarsening} (a) shows a simple OpenCL kernel where a thread operates on a work-item of the one-dimensional input array,
\texttt{in}, at a time. The work-item to be operated on is specified by the value returned from the OpenCL \texttt{get\_global\_id()} API.
Figure~\ref{fig:coarsening} (b) shows the transformed code after applying a thread coarsen factor of two, where each thread processes two
elements of the input array.

Thread coarsening can improve performance through increasing instruction-level parallelism~\cite{Volkov:2008:BGT:1413370.1413402}, reducing
the number of memory-access operations~\cite{Yang:2012:UOC:2207222.2207225}, and eliminating redundant computation when the same value is
computed in every work-item. However, it can also have several negative side-effects such as reducing the total amount of parallelism and
increasing the register pressure, which can lead to slowdown performance. Determining when and how to apply thread coarsening is
non-trivial, because the best coarsening factor depends on the target program and the hardware architecture that the program runs
on~\cite{Volkov:2008:BGT:1413370.1413402,Magni:2014:AOT:2628071.2628087}.

Magni \etal show that machine learning techniques can be used to automatically construct effective thread-coarsening heuristics across GPU
architectures~\cite{Magni:2014:AOT:2628071.2628087}. Their approach considers six coarsening factors, ($1, 2, 4, 8, 16, 32$). The goal is
to develop a machine learning based model to decide whether an OpenCL kernel should be coarsened on a specific GPU architecture and if so
what is the best coarsening factor. Among many machine learning algorithms, they chose to use an artificial neural network to
model\footnote{In fact, Magni \etal employed a hierarchical approach consisting of multiple artificial neural
networks~\cite{Magni:2014:AOT:2628071.2628087}. However, these networks are trained using the same process.} the problem. Construing such a
model follows the classical 3-step supervised learning process, which is depicted in Figure~\ref{fig:mlflow} and described in more details
as follows.

\begin{table}[t]
\caption{Candidate code features used in \cite{Magni:2014:AOT:2628071.2628087}.}
\label{tbl:rawfeatures}
\scriptsize
\begin{tabularx}{\columnwidth}{X||X}

\toprule

\textbf{Feature Description} & \textbf{Feature Description}\\

\midrule
\rowcolor{Gray} \# Basic Blocks & \# Branches \\

\# Divergent Instr. & \# Instrs. in Divergent Regions \\
\rowcolor{Gray}  (\# instr. in Divergent regions)/(\# total instr.) & \# Divergent regions \\
 \# Instrs & \# Floating point instr. \\
\rowcolor{Gray}  Avg. ILP per basic block & (\# integer instr.) / (\# floating point instr.) \\
\# integer instr. & \# Math built-in func.\\
\rowcolor{Gray} Avg. MLP per basic block & \# loads \\
\# stores & \# loads that are independent of the coarsening direction \\
\rowcolor{Gray}  \# barriers & \\

\bottomrule
\end{tabularx}
\end{table}

\paragraph{Feature engineering}
To describe the input OpenCL kernel, Magni \etal use static code features extracted from the compiler's intermediate representation.
Specifically, they developed a compiler-based tool to obtain the feature values from the program's LLVM
bitcode~\cite{Lattner:2004:LCF:977395.977673}. They started from 17 candidate features. These include things like the number of and types
of instructions and memory level parallelism (MLP)  within an OpenCL kernel. Table~\ref{tbl:rawfeatures} gives the list of candidate
features used in ~\cite{Magni:2014:AOT:2628071.2628087}. Typically, candidate features can be chosen based on developers' intuitions,
suggestions from prior works, or a combination of both. After choosing the candidate features, a statistical method called Principal
Component Analysis (see also Section~\ref{sec:unl}) is applied to map the 17 candidate features into 7 aggregated features, so that each
aggregated feature is a linear combination of the original features. This technique is known as ``\emph{feature dimension reduction}" which
is discussed at Section~\ref{sec:dimred}. Dimension reduction helps eliminating  redundant information among candidate features, allowing
the learning algorithm to perform more effectively.

\paragraph{Learning the model}
For the work presented in~\cite{Magni:2014:AOT:2628071.2628087}, 16 OpenCL benchmarks were used to generate training data. To find out
which of the six coarsening factors performs best for a given OpenCL kernel on a specific GPU architecture, we can apply each of the six
factors to an OpenCL kernel and records its execution time. Since the optimal thread-coarsening factor varies across hardware
architectures, this process needs to repeat for each target architecture. In addition to finding the best-performing coarsening factor,
Magni \etal also extracted the aggregated feature values for each kernel. Applying these two steps on the training benchmarks results in a
training dataset where each training example is composed of the optimal coarsening factor and feature values for a training kernel. The
training examples are then fed into a learning algorithm which tries to find a set of model parameters (or weights) so that overall
prediction error on the training examples can be minimised. The output of the learning algorithm is  an artificial neural network model
where its weights are determined from the training data.

\paragraph{Deployment}
The learned model can then be used to predict the optimal coarsening factor for \emph{unseen} OpenCL programs. To do so, static source code
features are first extracted from the target OpenCL kernel; the extracted feature values are then fed into the model which decides whether
to coarsen or not and which coarsening factor should use. The technique proposed in~\cite{Magni:2014:AOT:2628071.2628087} achieves an
average speedup between 1.11x and 1.33x across four GPU architectures and does not lead to degraded performance on a single benchmark.

%% file: methodology.tex
\section{Methodology \label{sec:methodology}}
\begin{figure}[t!]
	\centering %
	\subfloat[Use a cost function to guide compiler decisions]{%
     \includegraphics[width=0.5\textwidth]{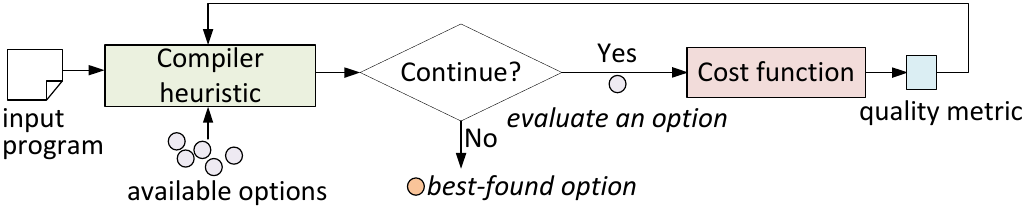} \label{subfig:searchcf}
	}\\
    \subfloat[Use a model to directly predict the decision]{%
     \includegraphics[width=0.45\textwidth]{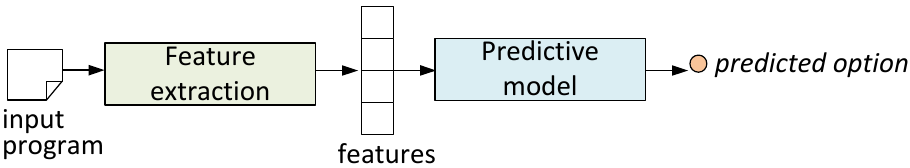} \label{subfig:directp}
	}
	\caption{There are in general two approaches to determine the optimal compiler decision using machine learning. The
first one is to learn a cost or priority function to be used as a proxy to select the best-performing option (a). The
second one is to learn a predictive model to directly predict the best option. }
	\label{fig:twoapproaches}
\end{figure}
One of the key challenges for compilation is to select the right code transformation, or sequence of transformations for a given program.
This requires effectively evaluating the quality of a possible compilation option e.g. how will a code transformation affect eventual
performance.

A naive approach is to exhaustively apply each legal transformation option and then profile the program to collect the relevant performance
metric.  Given that many compiler problems have a massive number of options, exhaustive search and profiling is infeasible, prohibiting the
use of this approach at scale. This search based approach to compiler optimisation is known as iterative compilation
\cite{bodin1998iterative,knijnenburg2003combined} or auto-tuning \cite{datta2008stencil,FFTW05}. Many techniques have been proposed to
reduce the cost of searching a large space \cite{Agakov:2006:UML:1121992.1122412,Nobre:2016:GIC:2907950.2907959}. In certain cases, the
overhead is justifiable if the program in question is to be used many times e.g. in a deeply embedded device. However, its main limitation
remains: it only finds a good optimisation for one program and does not generalise into a compiler heuristic.

There are two main approaches for solving the problem of scalably selecting compiler options that work across programs. A high level
comparison of both approaches is given in Figure~\ref{fig:twoapproaches}. The first strategy attempts to develop a \emph{cost} (or
\emph{priority}) function to be used as a proxy to estimate the quality of a potential compiler decision, without relying on extensive
profiling. The second strategy is to directly predict the best-performing option.

\subsection{Building a cost function \label{sec:costf}}
Many compiler heuristics rely on a cost function to estimate the quality of a compiler option. Depending on the optimisation goal, the
quality metric can be execution time, the code size, or energy consumption etc.  Using a cost function, a compiler can evaluate a range of
possible options to choose the best one, without needing to compile and profile the program with each option.


\subsubsection{The problem of hand-crafted heuristics}
 Traditionally, a compiler cost function is manually crafted. For example, a heuristic of function inlining adds up a number of
relevant metrics, such as the number of instructions of the target function to be inlined, the callee and stack size after inlining, and
compare the resulted value against a pre-defined threshold to determine if it is profitable to inline a
function~\cite{leupers1999function}. Here, the importance or weights for metrics and the threshold are determined by compiler developers
based on their experience or via ``trail-and-error". Because the efforts involved in tuning the cost function is so expensive, many
compilers simply use ``one-size-fits-all" cost function for inlining. However, such a strategy is ineffective.  For examples, Cooper \etal
show that a ``one-size-fits-all" strategy for inlining often delivers poor performance~\cite{Cooper:2008:ASI:1788374.1788381}; other
studies also show that that the optimal thresholds to use to determine when to inline changes from one program to the
other~\cite{Simon:2013:ACI:2495258.2495914,zhao2004inline}.

Hand-crafted cost functions are widely used in compilers. Other examples include the work conducted by Wagner
\etal~\cite{Wagner:1994:ASE:178243.178251} and Tiwari \etal\cite{629825}. The former combines a Markov model and a human-derived heuristic
to statically estimate the execution frequency of code regions (such as function innovation counts). The later calculates the energy
consumption of an applicaiton by assigning a weight to each instruction type. The efficiency of these approaches highly depend on the
accuracy of the estimations given by the manually tuned heuristic.

The problem of relying on a hand-tuned heuristic is that the cost and benefit of a compiler optimisation often depends on the underlying
hardware; while hand-crafted cost functions could be effective, manually developing one can take months or years on a single architecture.
This means that tuning the compiler for each new released processor is hard and is often infeasible due to the drastic efforts involved.
Because cost functions are important and manually tuning a good function is difficult for each individual architecture, researchers have
investigated ways to use machine learning to automate this process.

In the next subsection, we review a range of previous studies on using machine learning to tune cost functions for performance and energy
consumption -- many of which can be applied to other optimisation targets such as the code size~\cite{Cooper1999} or a trade-off between
energy and runtime.

\begin{figure*}[t!]
	\centering %
	\subfloat[An example cost function in \cite{metaopt}]{%
     \includegraphics[width=0.25\textwidth]{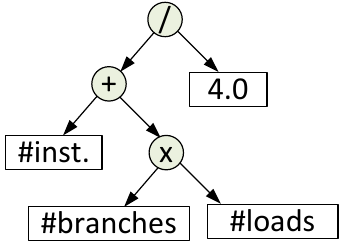} \label{subfig:costf:example}
	}
    \subfloat[A simple view of the generic programming technique in \cite{metaopt}]{%
     \includegraphics[width=0.8\textwidth]{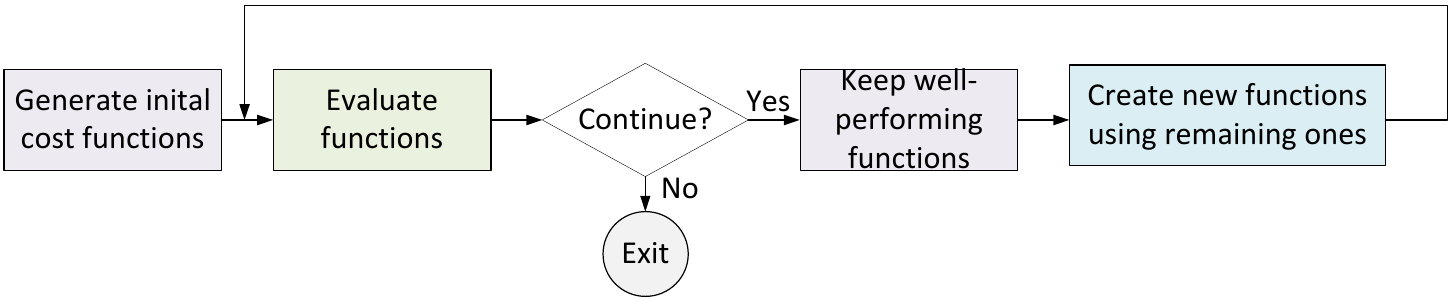} \label{subfig:gpview}
	}
	\caption{A simple view of the genetic programming (\GP) approach presented at \cite{metaopt} for tuning compiler
cost functions. Each candidate cost function is represented as an expression tree (a). The workflow of the \GP algorithm
is presented at (b). }
	\label{fig:gpforcostf}
\end{figure*}

\subsubsection{Cost functions for performance}


The Meta Optimization framework~\cite{metaopt} uses genetic programming (\GP) to search for a cost function, $y \leftarrow f(x)$, which
takes in a feature vector, $x$, and produces a real-valued priority, $y$. Figure~\ref{fig:gpforcostf} depicts the workflow of the
framework. This approach is evaluated on a number of compiler problems, including hyperblock formation\footnote{Hyperblock formation
combines basic blocks from multiple control paths to form a predicated, larger code block to expose instruction level parallelism.},
register allocation, and data prefetching, showing that machine learned cost functions outperform human-crafted ones. A similar approach is
employed by Cavazos \etal  find cost functions for performance and compilation overhead for a Java just-in-time
compiler~\cite{Cavazos:2005:ATI:1105760.1105779}. The COLE compiler~\cite{Hoste:2008:CCO:1356058.1356080} uses a variance of the \GP
algorithm called Strength Pareto Evolutionary Algorithm2 (\texttt{SPEA2})~\cite{Kim2004} to learn cost functions to balance multiple
objectives (such as program runtime, compilation overhead and code size). In Section~\ref{sec:rl}, we describe the working mechanism of
\GP-like search algorithms.

Another approach to tune the cost functions is to predict the execution time or speedup of the target program. The Qilin
compiler~\cite{Luk:2009:QEP:1669112.1669121} follows such an approach. It uses curve fitting algorithms to estimate the runtime for
executing the target program of a given input size on the CPU and the GPU. The compiler then uses this information to determine the optimal
loop iteration partition across the CPU and the GPU. The Qilin compiler relies on an application-specific function which is built on a per
program base using reference inputs. The curve fitting (or regression -- see also Section~\ref{sec:models}) model employed by the Qilin
compiler can model with continuous values, making it suitable for estimating runtime and speedup.  In \cite{yw14}, this approach is
extended, which developed a relative
 predictor that predicts whether an unseen predictor will improve significantly on a GPU relative to a CPU. This is used for runtime scheduling of OpenCL jobs.

The early work conduced by Brewer proposed a regression-based model to predict the execution of a data layout scheme for parallelization,
by considering three parameters~\cite{Brewer:1995:HOV:209936.209946}. Using the model, their approach can select the optimal layout for
over 99\% of the time for a Partial Differential Equations (PDE) solver across four evaluation platforms. Other previous works also use
curve fitting algorithms to build a cost function to estimate the speedup or runtime of sequential~\cite{4145110,
Lee:2006:AER:1168857.1168881,Park:2011:PMP:2190025.2190059}, OpenMP
~\cite{Curtis-Maury:2008:PMM:1454115.1454151,Singh2010,Wang:2009:MPM:1504176.1504189}, and more recently for deep learning
applications~\cite{Kang:2017:NCI:3037697.3037698}.



\subsubsection{Cost functions for energy consumption}
In addition to performance, there is an extensive body of work investigates ways to build energy models for software optimisation and
hardware architecture design. As power or energy readings are continuous real values, most of the prior work on power modelling use
regression-based approaches.

Linear regression is a widely used technique for energy modeling. Benini \etal developed a linear regression-based model to estimate power
consumption at the instruction level~\cite{Benini:1998:RMB:1275815.1275817}. The framework presented by Rethinagiri \etal.~\cite{6089692}
uses parameterised formulas to estimate power consumption of embedded systems. The parameters of the formulas are determined by applying a
regression-based algorithm to reference data obtained with hand-crafted assembly code and power measurements. In a more recent work,
Sch\"{u}rmans \etal also adopt a regression-based method for power modelling~\cite{Schurmans:2016:FEP:3008024.2987375}, but the weights of
the regression model are determined using standard benchmarks instead of hand-written assembly programs.

Other works employ the artificial neural network (\ANN) to automatically construct power models.
Curtis-Maury \etal develop an \ANN-based  model to predict the power consumption of OpenMP programs on multi-core
systems~\cite{4629274}. The inputs to the model are  hardware performance counter values such as the cache miss rate,
and the output is the estimated power consumption. Su \etal adopt a similar approach by developing an \ANN predictor
to estimate the runtime and power consumption for mapping OpenMP programs on Non-Uniform Memory Access (\NUMA)
multi-cores. This approach is also based on runtime profiling of the target program, but it explicitly considers
\NUMA-specific information like
local and remote memory accesses per cycle.

\subsection{Directly predict the best option}

While a cost function is useful for evaluating the quality of compiler options,
the overhead involved in searching for the optimal option may still be prohibitive. For this reason, researchers have investigated ways to
directly predict the best compiler decision using machine learning for relatively small compilation problems.

Monsifrot \etal pioneered the use of machine learning to predict the optimal compiler decision~\cite{monsifrot2002machine}. This work
developed a decision tree based approach to determine whether it is beneficial to unroll a loop based on information such as the number of
statements and arithmetic operations of the loop. Their approach makes a binary decision on whether to unroll a loop but not how many times
the loop should be unrolled. Later, Stephenson and Amarasinghe advanced \cite{monsifrot2002machine} by directly predicting the loop unroll
factor~\cite{Stephenson:2005:PUF:1048922.1048981} by considering eight unroll factors, ($1,2,\dots,8$). They formulated the problem as a
multi-class classification problem (i.e. each loop unroll factor is a class). They used over 2,500 loops from 72 benchmarks to train two
machine learning models (a nearest neighbor and a support vector machines model) to predict the loop unroll factor for unseen loops. Using
a richer set of features than \cite{monsifrot2002machine}, their techniques correctly predict the unroll factor for 65\% of the testing
loops, leading to on average, a 5\% improvement for the SPEC 2000 benchmark suite.

For sequential programs, there is extensive work in  predicting the best compiler flags~\cite{Cavazos:2007:RSG:1251974.1252540,
Cavazos:2006:MDC:1167473.1167492},  code transformation options
~\cite{Dubach:2007:FCO:1242531.1242553},
or  tile size for loops~\cite{Yuki:2010:ACT:1772954.1772982, 6406709}.
This level of interest is possibly due to the restricted nature of the problem, allowing easy experimentation
and comparision against prior work.


Directly predicting the optimal option for parallel programs is harder than doing it for sequential programs, due to the complex
interactions between the parallel programs and the underlying parallel architectures. Nonetheless, there are works on predicting the
optimal number of threads to use to run an OpenMP program~\cite{Wang:2009:MPM:1504176.1504189,Moore2014}, the best parameters to used to
compile a CUDA programs for a given input~\cite{5160988} the thread coarsening parameters for OpenCL programs for
GPUs~\cite{Magni:2014:AOT:2628071.2628087}. These papers show that supervised machine learning can be a powerful tool for modelling
 problems with a relatively small number of optimisation options.

%
%

%% file: models
\section{Machine Learning Models \label{sec:models}}
In this section, we review the wide range of machine learning models used for compiler optimisation.

There are two major subdivisions of machine learning techniques that have previously been used in compiler optimisations: supervised and
unsupervised learning. Using supervised machine learning, a predictive model is trained on empirical performance data (\emph{labelled
outputs}) and important quantifiable properties (features) of representative programs. The model learns the correlation between these
feature values and the optimisation decision that delivers the optimal (or nearly optimal) performance. The learned correlations are used
to predict the best optimization decisions for new programs. Depending on the nature of the outputs, the predictive model can be either a
\emph{regression} model for continuous outputs or a \emph{classification} model for discrete outputs.

In the other subdivision of machine learning, termed \emph{unsupervised learning}, the input to the learning algorithm is a set of input
values merely -- there is no labelled output. One form of unsupervised learning is \emph{clustering} which groups the input data items into
several subsets. For example, SimPoint~\cite{Perelman:2003:USA:781027.781076}, a simulation technique, uses clustering to pick represent
program execution points for program simulation. It does so by first dividing a set of program runtime information into groups (or
clusters), such that points within each cluster are similar to each other in terms of program structures (loops, memory usages etc.); it
then chooses a few points of each cluster to represent all the simulation points within that group without losing much information.

There are also techniques that sit at the boundary of supervised and unsupervised learning. These techniques refine the knowledge gathered
during offline learning or previous runs using empirical observations obtained during deployment. We review such techniques in
Section~\ref{sec:rl}. This sections concludes with a discussion of the relative merits of different modelling approaches for compiler
optimisation.

\subsection{Supervised Learning \label{sec:sl}}
\subsubsection{Regression}
A widely used supervised learning technique is called \emph{regression}. This technique has been used in various tasks, such as predicting
the program execution time input~\cite{Luk:2009:QEP:1669112.1669121} or speedup~\cite{yw14} for a given input, or estimating the tail
latency for parallel workloads~\cite{Zhang:2016:TAS:3001136.3001186}.

\begin{figure}[tb]
\begin{center}
    \includegraphics[width=0.45\textwidth]{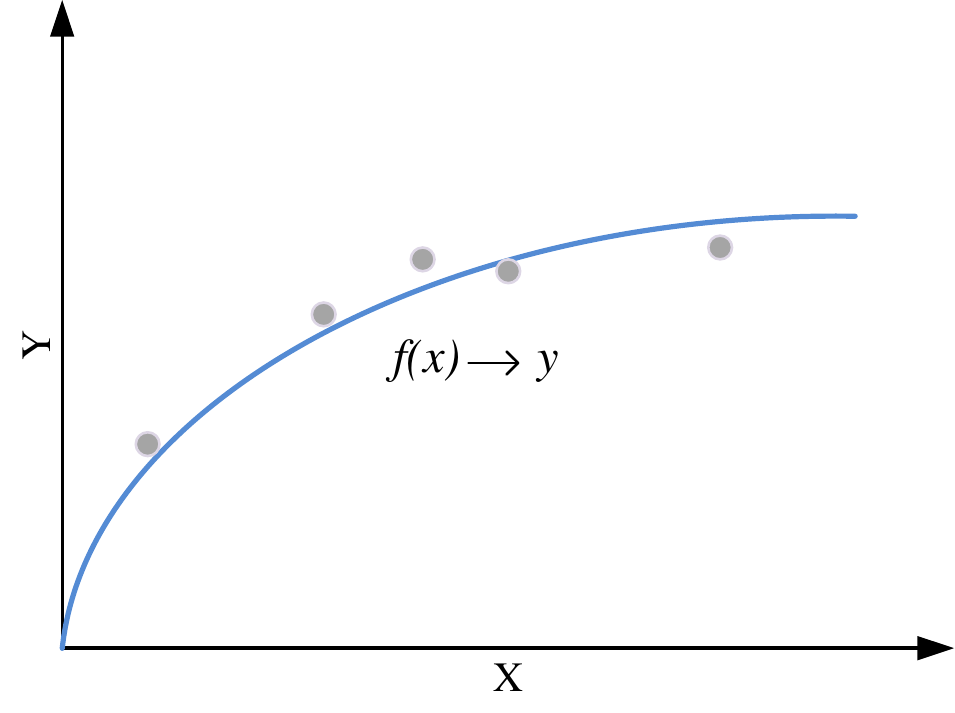}
  \end{center}
\caption{A simple regression-based curve-fitting example. There are five training examples in this case. A function, $f$, is trained with the training data, which maps the input $x$ to the output $y$. The trained function can predict the output of an unseen $x$.}
\label{fig:regression}
\end{figure}

Regression is essentially curve-fitting. As an example, consider Figure~\ref{fig:regression} where a regression model is learned from five
data points. The model takes in a program input size, $X$, and predicts the execution time of the program, $Y$. Adhering to supervised
learning nomenclature, the set of five known data points is the training data set and each of the five points that comprise the training
data is called a training example. Each training example, $(x_i, y_i)$, is defined by a feature vector (i.e. the input size in our case),
$x_i$, and a desired output (i.e. the program execution time in our case), $y_i$. Learning in this context is understood as discovering the
relation between the inputs ($x_i$) and the outputs ($y_i$) so that the predictive model can be used to make predictions for any new,
unseen input features in the problem domain. Once the function, $f$, is in place, one can use it to make a prediction by taking in a new
input feature vector, $x$. The prediction, $y$, is the value of the curve that the new input feature vector, $x$, corresponds to.

There are a range of machine learning techniques can be used for curve-fitting. These include the simple linear regression model and more
advanced models like support vector machines (\SVMs) and neural networks (\NN). Linear regression is effective when the input (i.e. feature
vectors) and output (i.e. labels) have a strong linear relation. \SVM and \NN can model both linear and non-linear relations, but typically
require more training examples to learn an effective model when compared with linear regression.

Table~\ref{tbl:regression} gives some examples of regression techniques that have been used in prior work for code optimisation and the
problem to be modelled.

\begin{table}[t]
\caption{Regression techniques used in prior works.}
\label{tbl:regression}
\begin{tabular}{lll}
\toprule
Modelling Technique & Application & References\\
\midrule
Linear Regression & Exec. Time Estimation & \cite{Lee:2007:MIL:1229428.1229479,Luk:2009:QEP:1669112.1669121,Park:2011:PMP:2190025.2190059} \\
Linear Regression & Perf. \& Power Prediction & \cite{Curtis-Maury:2006:OPA:1183401.1183426,6957246,Berral:2010:TES:1791314.1791349} \\
Artificial Neural Networks & Exec. Time Estimation & \cite{Lee:2007:MIL:1229428.1229479,Wang:2009:MPM:1504176.1504189,yw14} \\

\bottomrule
\end{tabular}
\end{table}

\subsubsection{Classification}
Supervised classification is another technique that has been widely used in prior work of machine learning based code optimisation.  This
technique takes in a feature vector and predicts which of a set of classes the feature vector is associated with. For example,
classification can be used to predict which of a set of unroll factors should be used for a given loop, by taking in a feature vector that
describes the characteristics of the target loop (see also Section~\ref{sec:motivation}).

The k-nearest neighbour (\KNN) algorithm is a simple yet effective classification technique. It finds the $k$ closet training examples to
the input instance (or program) on the feature space. The closeness (or distance) is often evaluated using the Euclidean distance, but
other metrics can also be used.  This technique has been used to predict the optimal optimisation parameters in prior
works~\cite{Stephenson:2005:PUF:1048922.1048981,DelVento:2012:POS:2185475.2185477, micolet2016machine}. It works by first predicting which
of the training programs are closet (i.e. nearest neighbours) to the incoming program on the feature space; it then uses the optimal
parameters (which are found during training time) of the nearest neighbours as the prediction output. While it is effective on small
problems, \KNN also has two main drawbacks. Firstly, it must compute the distance between the input and all training data at each
prediction. This can be slow if there is a large number of training programs to be considered. Secondly, the algorithm itself does not
learn from the training data; instead, it simply selects the $k$ nearest neighbours. This means that the algorithm is not robust to noisy
training data and could choose an ill-suited training program as the prediction.

As an alternative, the decision tree has been used in prior works for a range of optimisation problems. These include choosing the parallel
strategy for loop parallelization~\cite{Yu:2000:ARP:335231.335238}, determining the loop unroll factor~\cite{monsifrot2002machine,
Leather:2009:AFG:1545006.1545059}, deciding the profitability of using GPU acceleration~\cite{6494993}, and selecting the optimal algorithm
implementation~\cite{ Ding:2015:AAC:2737924.2737969}. The advantage of a decision tree is that the learned model is interpretable and can
be easily visualised (see Figure~\ref {fig:unrolling}). This enables users to understand why a particular decision is made by following the
path from the root node to a leaf decision node.

Decision trees make the assumption that the feature space is convex i.e. it can be divided up using hyperplanes  into different regions
each of which belongs to a different category. This restriction is often appropriate in practice. However, a significant  drawback of using
a single decision tree is that the model can over-fit due to outliers in the training data (see also Section~\ref{sec:model:discuss}).
Random forests~\cite{Ho:1995:RDF:844379.844681} have therefore been proposed to alleviate the problem of over fitting. Random forests works
by constructing multiple decision trees at training time. The prediction of each tree depends on the values of a random vector sampled
independently on the feature value. In this way, each tree is randomly forced to be insensitive to some feature dimensions. To make a
prediction, random forests then aggregate the outcomes of individual trees to form an overall prediction. It has been employed to determine
whether to inline a function or not~\cite{lokuciejewski2009automatic}, delivering better performance than a single-model-based approach. It
is to note that random forests can also be used for regression tasks. For instances, it has  been used to model energy consumption of
OpenMP~\cite{7284455} and CUDA~\cite{rejitha2017energy} programs.

Logical regression is a variation of linear regression but is often used for classification. It takes in the feature vector and calculate
the probability of some outcome. For example, Cavazos and O'Boyle used logical regression to determine the optimization level of Jike RVM.
Like decision trees, logical regression also assumes that the feature values and the prediction has a linear relation.

More advanced models such as \SVM classification has been used for various compiler optimisation
tasks~\cite{Wang:2009:MPM:1504176.1504189,Taylor2017AdaptiveOF}. \SVMs use kernel functions to compute the similarity of feature vectors.
The radial basis function is commonly used in support vector machine classification because it model both linear and non-linear problems.
However, other kernels can also be used. Other machine learning techniques such as Kernel Canonical Correlation Analysis and naive Bayes
have also been used in prior works to predict stencil program configurations~\cite{Ganapathi:2009:CML:1855591.1855592}  or detect parallel
patterns~\cite{Deniz2016}.

\subsection{Unsupervised Learning}
\begin{figure}
  \centering
  \includegraphics[width=0.35\textwidth]{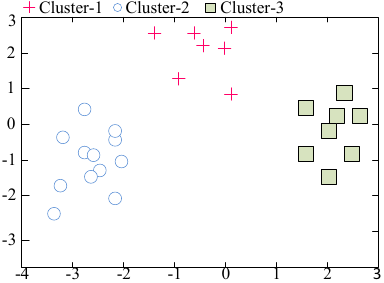}\\
  \caption{Using k-means to group data points into three clusters. In this example, we group the data points into three clusters
  on a 2-d feature space.}\label{fig:kmeans}
\end{figure}

Unlike supervised learning models which learn a correlation from the input feature values to the corresponding outputs, unsupervised
learning models only take it the input data (e.g. the feature values). This technique is often used to model the underlying structure of
distribution of the data.

Clustering is a classical unsupervised learning problem. The k-means clustering algorithm~\cite{ macqueen1967some} groups the input data
into $k$ clusters. For example, in Figure~\ref{fig:kmeans}, a k-means algorithm is used to group data points into three clusters on a
2-dimensional feature space. The algorithm works by grouping data points that are close to each other on the feature space into a cluster.
K-means is used to characterize program behaviour~\cite{Sherwood:2002:ACL:605397.605403,Perelman:2003:USA:781027.781076}. It does so by
clustering program execution into phase groups, so that we can use a few samples of a group to represent the entire program phases within a
group. K-means is also used in the work presented in \cite{Wang:2010:PSP:1854273.1854313} to summarise the code structures of parallel
programs that benefit from similar optimisation strategies.

{\em MOB should talk about PCA and maybe auto-encoders as these are unsupervised techniques -primarily used for feature normalisation etc.
Maybe a forward reference to feature section would be sufficient??}

\subsection{Reinforcement Learning \label{sec:rl}}
Reinforcement learning (\RL),  sometimes called ``learning from interaction",  can seen as an online learning algorithm.  The algorithm
tries to learn how to maximise the rewards (or performance) itself. In other words, the algorithm needs to learn, for a given input, what
is the correct output or decision to take. This is different from supervised learning where the correct input/output pairs are presented.

\begin{figure}
  \centering
  \includegraphics[width=0.35\textwidth]{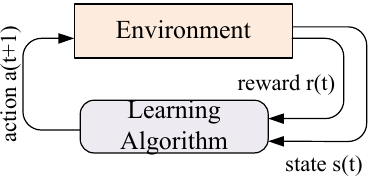}\\
  \caption{The working mechanism of reinforcement learning}\label{fig:rl}
\end{figure}

Figure~\ref{fig:rl} illustrates the working mechanism of \RL. Here the learning algorithm interacts with its environment over a discrete
set of time steps. At each step, the algorithm evaluate the current \emph{state} of its environment, and executes an \emph{action}. The
action leads to a change in the state of the environment (which the algorithm can evaluate in the next time step), and produces an
immediate \emph{reward}. For examples, in a multi-tasking environment, a state could be the CPU contention and which processor cores are
idle, an action could be where to place a process, and a reward could be the overall system throughput. The goal of \RL is to maximize the
long-term cumulative reward by learning an optimal strategy to map states to actions.

\RL is particularly suitable for modelling problems that have an evolving natural, such as dynamic task scheduling, where the optimal
outcome is achieved through a series of actions. \RL has been used in prior works to schedule RAM memory traffics~\cite{ipek2008self},
selecting software component configurations at runtime~\cite{porter2016rex}, and configure virtual
machines~\cite{Rao:2009:VRL:1555228.1555263}.

\RL is an intuitive and comprehensive solution for autonomous decision making. But its performance depends on the effectiveness of the
value function, which estimates the immediate reward. An optimal value function should lead to the greatest cumulative reward in the longer
term. For many problems, it is difficult to find an effective value function or policy, because the function needs to foresee the impact of
an action in the future. In recent years, deep learning techniques have been used in conjunct with \RL to learn a value function. The
combined technique is able to solve some problems that were deem impossible in the past~\cite{DBLP:journals/corr/Li17b}. However, how to
combine deep learning with \RL to solve compilation and code optimisation problems remains an open question.

{\em MOB  first real mention of deep learingin. Should we mention this in the supervised models section??}

{\em MOB maybe add Mutali LCPC -this uses a Q function form RL}

\subsection{Discussions \label{sec:model:discuss}}

What model is best, is the \$64,000  question.  The answer is : it depends. More sophisticated techniques may provide greater accuracy but
they require large amounts of labelled training data - a real problem in compiler optimisation. Techniques like linear regression and
decision trees require less training data compared to more advanced models like \SVM and \ANN. Simple models typically work well when the
prediction problem can be described using a feature vector that has a small number of dimensions, and when the feature vector and the
prediction is linearly correlated. More advanced techniques like \SVM and \ANN can model both linear and non-linear problems on a higher
dimensional feature space, but they often require more training data to learn an effective model. Furthermore, the performance of \SVM and
\ANN also highly depends the hyper-parameters used to train the model. The optimal hyper-parameter values can be chosen by performing
cross-validation on the training data. However, how to select the right parameters to avoid over-fitting while achieving a good prediction
accuracy remains an open problem.

Choosing which modelling technique to use is non-trivial. This is because the choice of model depends on a number of factors: the
prediction problem (e.g. regression or classification), the set of features to use, the available training examples, the training and
prediction overhead, etc. In prior works, the choice of modelling technique is largely relied on developer experience and empirical
results. Many of the studies do not fully justify the choice of the model, although some do compare the performance of alternate
techniques.

One technique that has seen little investigation is the use of Gaussian Processes \cite{}. Before the recent widespread interest in deep
neural networks, these were a highly popular method in many areas of machine learning \cite{}. They are particular powerful when the amount
of training data is sparse and expensive to collect. They also automatically give a confidence interval with any decision. This allows the
compiler writer to trade off risk vs reward depending on application scenario,

Using a single model has a significant drawback in practice. This is because a one size-fits-all model is unlikely to precisely capture
behaviors of diverse applications, and no matter how parameterized the model is, it is highly unlikely that a model developed today will
always be suited for tomorrow. To allow the model to adapt to the change of the computing environment and workloads, ensemble learning was
exploited in prior works~\cite{Ho:1995:RDF:844379.844681,Emani:2015:CDM:2737924.2737999,nguyen2017introduction}. The idea of ensemble
learning is to use multiple learning algorithms, where each algorithm is effective for particular problems, to obtain better predictive
performance than could be obtained from any of the constituent learning algorithm alone~\cite{polikar2006ensemble, rokach2010ensemble}.
Making a prediction using an ensemble typically requires more computational time than doing that using a single model, so ensembles can be
seen as a way to compensate for poor learning algorithms by performing extra computation. To reduce the overhead, fast algorithms such as
decision trees are commonly used in ensemble methods (e.g. Random Forests), although slower algorithms can benefit from ensemble techniques
as well.

%% file: feature.tex
\section{Feature Engineering \label{sec:featureengineering}}

\begin{table}[t]
\caption{Summary of features discussed in Section~\ref{sec:featureengineering}.}
\label{tbl:fsummary}
\begin{tabularx}{\columnwidth}{X X}
\toprule
\textbf{Feature} & \textbf{Description}\\
\midrule
\rowcolor{Gray} Static code features & Features gathered from source code or the compiler intermediate representations, such as instruction counts. See Section~\ref {sec:scf}.  \\

Tree and graph based features & Features extracted from the program graph, such as the number of nodes of different types. See Section~\ref {sec:tgf}.\\

\rowcolor{Gray} Dynamic features & Features obtained through dynamic profiling or during runtime execution, such as performance counter values. See Section~\ref {sec:dyf}. \\

Reaction-based features & Speedups or execution time obtained by profiling the target program under specific compiler settings. See Section~\ref {sec:rbf}.\\

\bottomrule
\end{tabularx}

\end{table}

\begin{table}[t]
\caption{Feature engineering techniques discussed in Section~\ref{sec:featureengineering}.}
\label{tbl:fegineering}
\begin{tabularx}{\columnwidth}{X X}
\toprule
\textbf{Problem} & \textbf{Techniques}\\
\midrule

\rowcolor{Gray} Feature selection & Pearson correlation coefficient, mutual information, regression analysis. See Section~\ref{sec:fesecl}.\\

Feature dimensionality reduction & Principal component analysis (\PCA), factor analysis, linear discriminant analysis, autoencoder. See Section~\ref {sec:dimred}.\\

\bottomrule
\end{tabularx}

\end{table}

Machine learning based code optimisation relies on having a set of high-quality features that  capture the important characteristics of the
target program. Given that there is an unbounded  number of potential features, finding the right set is a non-trivial task. In this
section, we review how previous  work chooses features, a task known as feature engineering. Tables~\ref{tbl:fsummary} and
\ref{tbl:fegineering} summarises the range of program features and feature engineering techniques discussed in this section, respectively.

\subsection{Feature representation}
Various forms of program features have been used in compiler-based machine learning. These include static code
structures~\cite{Jiang:2010:ESC:1772954.1772989} and runtime information such as system
load~\cite{Emani:2015:CDM:2737924.2737999,middleware17} and performance counters~\cite{Cavazos:2007:RSG:1251974.1252540}.

\subsubsection{Static code features \label{sec:scf}}
\begin{table}[t]
\caption{Example code features used in prior works.}
\label{tbl:codefexample}
\centering
\begin{tabularx}{\columnwidth}{X X}
\toprule
Description & Examples\\
\midrule
 \rowcolor{Gray} Arithmetic instructions & \#floating point instr., \#integer instr., \#method call instr. \\
Memory operations & \#load instr, \#store instr.\\
 \rowcolor{Gray} Branch instructions & \#conditional branch instr, \#unconditional branch instr\\
 loop information & \#loops, loop depth\\
 \rowcolor{Gray} parallel information & \#work threads, work group size\\
\bottomrule
\end{tabularx}
\end{table}

Static program features like the number and type of instructions are often used to describe a program. These features are typically
extracted from the compiler intermediate representations~\cite{Wang:2009:MPM:1504176.1504189,Simon:2013:ACI:2495258.2495914,
Stephenson:2005:PUF:1048922.1048981,Taylor2017AdaptiveOF} in order to avoid using information extracted from dead code.
Table~\ref{tbl:codefexample} gives some of the static code features that were used in previous studies. Raw code features are often used
together to create a combined feature. For example, one can divide the number of load instructions by the number of total instructions to
get the memory load ratio. An advantage of using static code features is that the features are readily available from the compiler
intermediate representation.

\subsubsection{Tree and graph based features \label{sec:tgf}}
Singer and Veloso represent the \FFT in a split tree~\cite{Singer:2000:LPP:645529.657967}. They extract from the tree a set of features, by
counting the number of nodes of various types and quantifying the shape of the tree. These tree-based features are then used to build a
neural network based cost function that predicts which of the two \FFT formulas runs faster. The cost function is used to search for the
best-performing transformation.

Park \emph{et al.} present a unique graph-based approach for feature representations~\cite{Park:2012:UGP:2259016.2259042}. They use a \SVM
where the kernel is based on a graph similarity metric. Their technique requires hand coded features at the basic block level, but
thereafter, graph similarity against each of the training programs takes the place of global features. Mailike shows that spatial based
information, i.e. how instructions are distributed within a program, extracted from the program's data flow graph could be useful features
for machine learning based compiler optimisation~\cite{5708970}. Nobre \emph{et al.} also exploit graph structures for code
generation~\cite{Nobre:2016:GIC:2907950.2907959}. Their approach targets the phase ordering problem. The order of compiler optimisation
passes is represented as a graph. Each node of the graph is an optimisation pass and connections between nodes are weighted in a way that
sub-sequences with higher aggregated weights are more likely to lead to faster runtime. The graph is automatically constructed and updated
using iterative compilation (where the target program is complied using different compiler passes with different orders). A design space
exploration algorithm is employed to drive the iterative compilation process.

\subsubsection{Dynamic Features \label{sec:dyf}}
\begin{figure}[t!]
  \centering
  \includegraphics[width=\columnwidth]{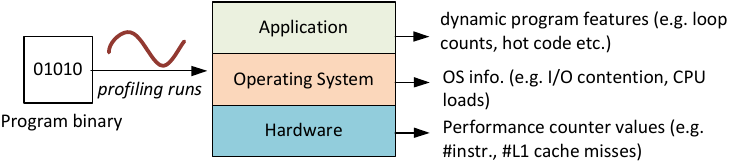}\\
  \caption{Dynamic features can be extracted from multiple layers of the computing environment.}\label{fig:dynf}
\end{figure}

While static code features are useful and can be extracted at static compile time (hence feature extraction has no runtime overhead), they
have drawbacks. For examples, static code features may contain information of code segments that rarely get executed, and such information
can confuse the machine learning model; some program information such as the loop bound depends on the program input, which can only
obtained during execution time; and static code features often may not precisely capture the application behaviour in the runtime
environment (such as resource contention and I/O behaviour) as such behaviour highly depends on the computing environment such as the
number of available processors and co-running workloads.

As illustrated in Figure~\ref{fig:dynf}, dynamic features can be extracted from multiple layers of the runtime environment. At the
application layer, we can obtain information like loop iteration counts the cannot be decided at compile time, dynamic control flows,
frequently executed code regions, etc. At the operating system level, we can observe the memory and I/O behaviour of the application as
well as CPU load and thread contention, etc. At the hardware level, we can use performance counters to track information like how many
instructions have been executed and of what types, and the number of cache loads/stores as well as branch misses, etc.

Hardware performance counter values like executed instruction counts and cache miss rate are therefore used to understand the application's
dynamic behaviours~\cite{Cavazos:2007:RSG:1251974.1252540,burtscher2012quantitative,luo2015fast}. These counters can capture low-level
program information such as data access patterns, branches and computational instructions. One of the advantage of performance counters is
that they capture how the target program behave on a specific hardware and avoid the irrelevant information that static code features may
bring in.  In addition to hardware performance counters, operating system level metrics like system load and I/O contention are also used
to model an application's behavior~\cite{yw14,middleware17}. Such information can be externally observed without instrumenting the code,
and can be obtain during off-line profiling or program execution time.

While effective, collecting dynamic information could incur prohibitively overhead and the collected information can be noisy due to
competing workloads and operating system scheduling~\cite{browne2000portable} or even subtle settings of the execution
environment~\cite{Mytkowicz:2009:PWD:1508244.1508275}. Another drawback of performance counters and dynamic features is that they can only
capture the application's past behavior. Therefore, if the application behaves significantly different in the future due to the change of
program phases or inputs, then the prediction will be drawn on an unreliable observation. As such, dynamic and static features are often
used in combination in prior works in order to build a robust model.



\subsection{Reaction based features \label{sec:rbf}}
\begin{figure}[t!]
	\centering %
	\subfloat[Static program feature based predictor]{%
     \includegraphics[width=0.5\textwidth]{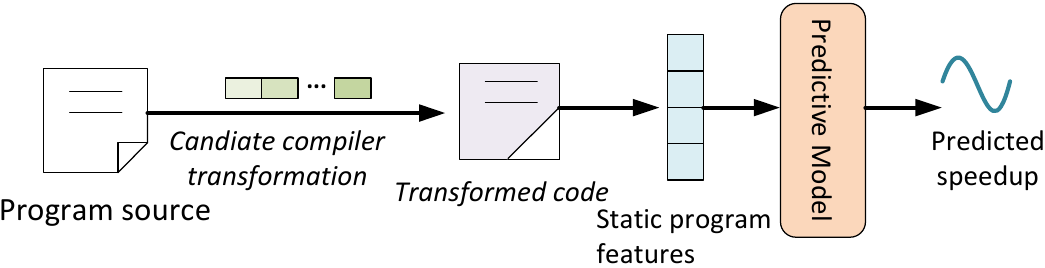} \label{subfig:searchcf}
	}\\
    \subfloat[Reaction based predictor]{%
     \includegraphics[width=0.5\textwidth]{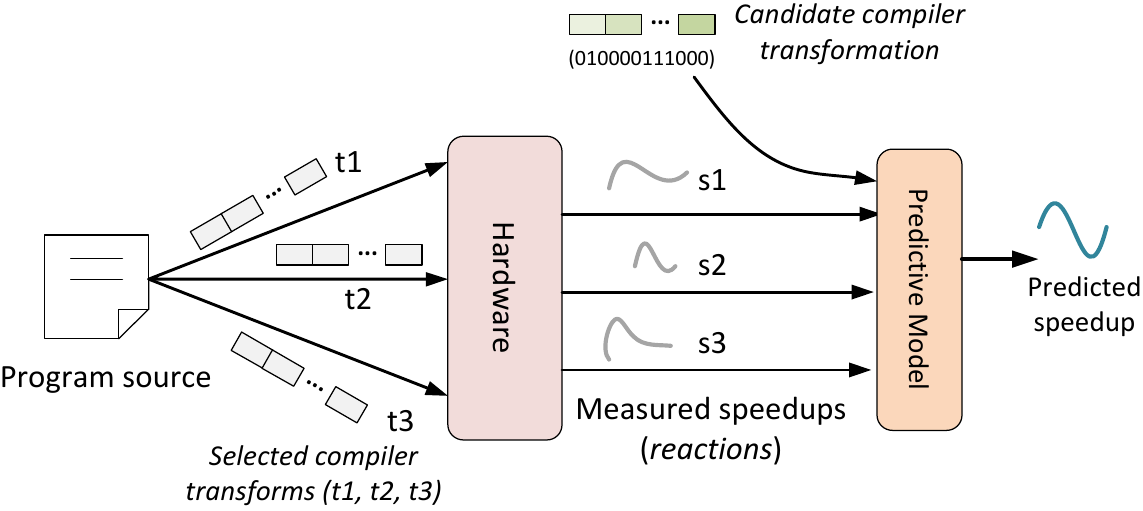} \label{subfig:directp}
	}
	\caption{Standard feature-based modelling (a) vs reaction-based modelling (b). Both models try to predict
the speedup for a given compiler transformation sequence. The program feature based predictor takes in static program features extracted
from the transformed program, while the reaction based model takes in the target transformation sequence and
the measured speedups of the target program, obtained by applying a number of carefully selected transformation sequences. Diagrams are reproduced
from~\cite{Cavazos:2006:APM:1176760.1176765}.}
	\label{fig:reactionbased}
\end{figure}

Cavazos \emph{et al.\ }present a reaction-based predictive model for software-hardware co-design~\cite{Cavazos:2006:APM:1176760.1176765}.
Their approach profiles the target program using several carefully selected compiler options to see how program runtime changes under these
options for a given micro-architecture setting. They then use the program ``reactions'' to predict the best available application speedup.
Figure~\ref{fig:reactionbased} illustrates the difference between a reaction-based model and a standard program feature based model. A
similar reaction-based approach is used in ~\cite{khan2007using} to predict  speedup and energy efficiency for an application that is
parallelised thread-level speculation (TLS) under a given micro-architectural configuration.
 Note that while a reaction-based approach does not use
static code features, developers must carefully select a few settings from a large number of candidate options for profiling, because
poorly chosen options can significantly affect the quality of the model.

\subsection{Automatic feature generation \label{sec:fe}}
As deriving good features is a time-consuming task, a few methods have been proposed to automatically generate features from the compiler's
intermediate representation (IR)~\cite{Namolaru:2010:PAS:1878921.1878951,Leather:2009:AFG:1545006.1545059}. The work of
\cite{Leather:2009:AFG:1545006.1545059} uses \GP to search for features, but required a huge grammar to be written, some 160kB in length.
Although much of this can be created from templates, selecting the right range of capabilities and search space bias is non trivial and up
to the expert. The work of \cite{Namolaru:2010:PAS:1878921.1878951} expresses the space of features via logic programming over relations
that represent information from the IRs. It greedily searches for expressions that represent good features. However, their approach relies
on expert selected relations, combinators and constraints to work. Both approaches closely tie the implementation of the predictive model
to the compiler IR, which means changes to the IR will require modifications to the model. Furthermore, the time spent in searching
features could be significant for these approaches.

The first work to employ neural network to extract features from program source code for compiler optimisation is conducted by Cummins
\emph{et al.} ~\cite{pact17}. Their system, namely DeepTune,  automatically abstracts and selects appropriate features from the raw source
code. Unlike prior work where the predictive model takes in a set of human-crafted features, program code is used directly in the training
data. Programs are fed through
a series of neural network based language models which learn how code correlates with the desired optimisation options (see also Figure~\ref{fig:dnn}).
Their work also shows that the properties of the raw code that are abstracted by the top layers of the neural networks are mostly
independent of the optimisation problem.
While promising, it is worth mentioning that dynamic information such as the program input size and performance counter values are often
essential for characterising the behaviour of the target program. Therefore, DeepTune does not completely remove human involvement for
feature engineering when static code features are insufficient for the optimisation problem.

\subsection{Feature selection and dimension reduction \label{sec:pca}}
Machine learning  uses features to capture the essential characteristics of a training example. Sometimes
we have too many features.
As the number of features increase so does the number of training examples needed to build an accurate model~\cite{Bishop:2006:PRM:1162264}.
Hence, we need to limit the dimension of the feature space
In compiler research, commonly, an initial large, high dimensional candidate
feature space is pruned via feature selection~\cite{Stephenson:2005:PUF:1048922.1048981}, or projected into a lower dimensional
space~\cite{Magni:2014:AOT:2628071.2628087}. In this subsection, we review a number of feature selection and dimension reduction methods.

\subsubsection{Feature selection \label{sec:fesecl}}
Feature selection requires understanding how does a particular feature affect the prediction accuracy. One of the simplest methods for
doing
this is  applying the Pearson correlation coefficient. This metric measures the linear correlation between two variables
and is used in
numerous works~\cite{hoste2006performance,Dubach:2007:FCO:1242531.1242553,Jiang:2010:ESC:1772954.1772989,Wang:2010:PSP:1854273.1854313} to
filter out redundant features by removing features that have a strong correlation with an already selected feature. It has also been used
to quantify the relation of the select features in regression.
One obvious drawback of using Pearson correlation as a feature ranking mechanism is that it is only sensitive to a linear relationship.

Another approach for correlation estimation is mutual
information~\cite{Cavazos:2006:APM:1176760.1176765,Rosenblum:2010:ECP:1806672.1806678}, which quantifies how much information of one
variable (or feature) can be obtained through another variable (feature). Like correlation coefficient, mutual information can be used to
remove redundant features. For example, if the information of feature, $x$, can be largely obtained through another existing feature, $y$,
feature $x$ can then be taken out from the feature set without losing much information on the reduced feature set.

Both correlation coefficient and mutual information evaluate each feature independently with respect to the prediction. A different
approach is to utilise regression analysis for feature ranking. The underlying principal of regression analysis is that if the prediction
is the outcome of regression model based on the features, then the most important features should have the highest weights (or
coefficients) in the model, while features uncorrelated with the output variables should have weights close to zero. For example, LASSO
(least absolute shrinkage and selection operator) regression analysis is used in \cite{7429329} to remove less useful features to build a
compiler-based model to predict performance. LASSO has also been used for feature selection to tune the compiler heuristics for the TRIPS
processor~\cite{Taylor:2010:ECH:2898607.2898878}.

In general, feature selection remains an open problem for machine learning, and researchers often follow a ``trail-and-error" approach to
test a range of methods and feature candidates. This makes automatic feature selection framework like FEAST~\cite{Ting2016} and
HERCULES~\cite{6957226} attractive. The former framework employs a range of existing feature selection methods to select useful candidate
features, while the latter searches for the most important static code features from a set of pre-defined patterns for loops.

\subsubsection{Feature dimensionality reduction\label{sec:dimred}}
While feature selection allows us to select the most important features, the resulted feature set can still be too large to train a good
model, especially when we only have a small number of training examples. By reducing the number of dimensions, the learning algorithm can
often perform more efficiently on a limited training dataset. Dimension reduction is also important for some machine learning algorithms
such as \KNN to avoid the effect of the curse of dimensionality~\cite{beyer1999nearest}.

\PCA is a well-established feature reduction technique~\cite{Fodor02asurvey}. It uses orthogonal linear transformations to reduce the
dimensionality of a set of variables i.e. features in our case.

\begin{figure}[t!]
   \centering
   \subfloat[Original feature space]{
        \includegraphics[width=0.22\textwidth]{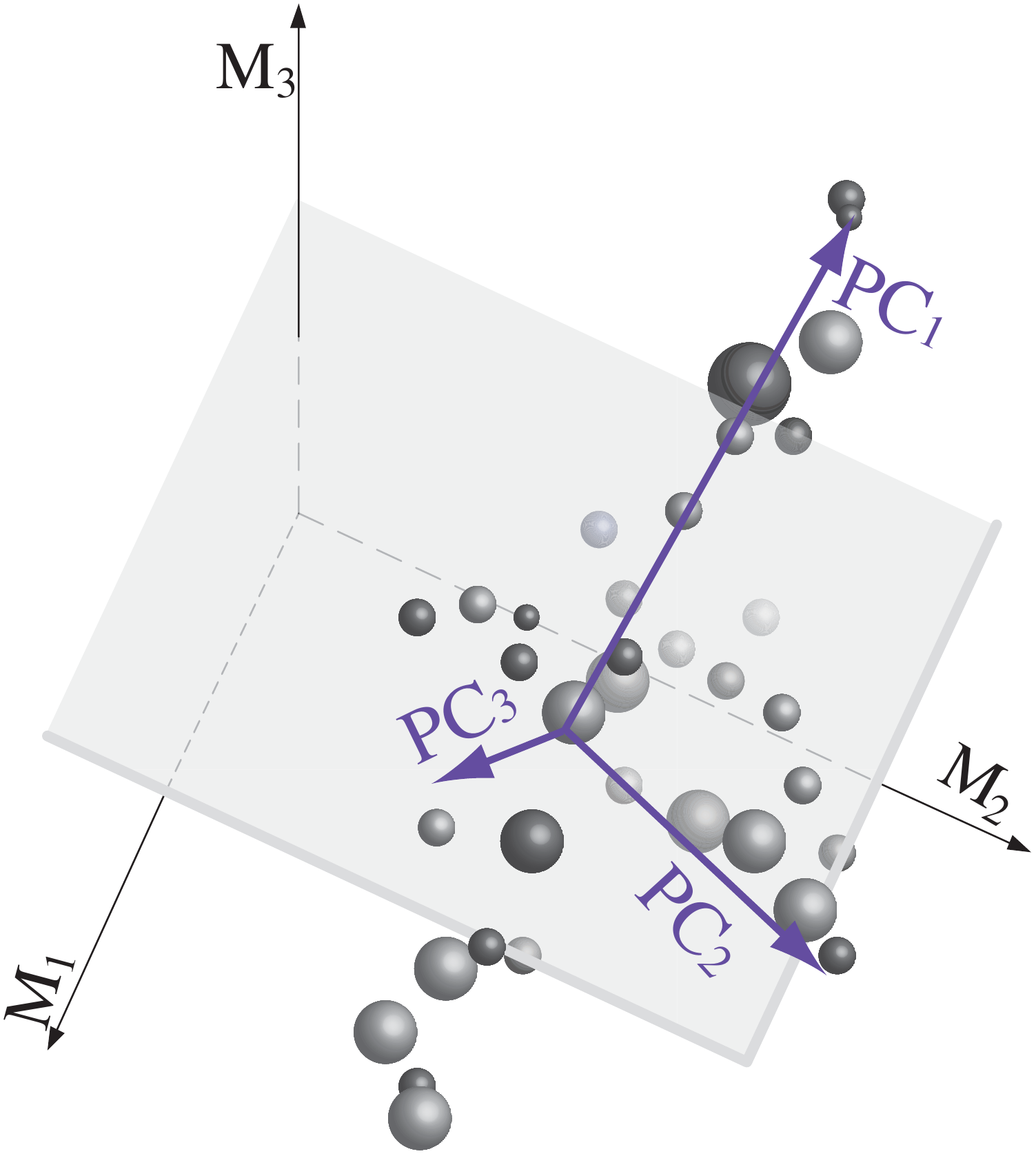}
  }
  \subfloat[Reduced feature space]{
        \includegraphics[width=0.22\textwidth]{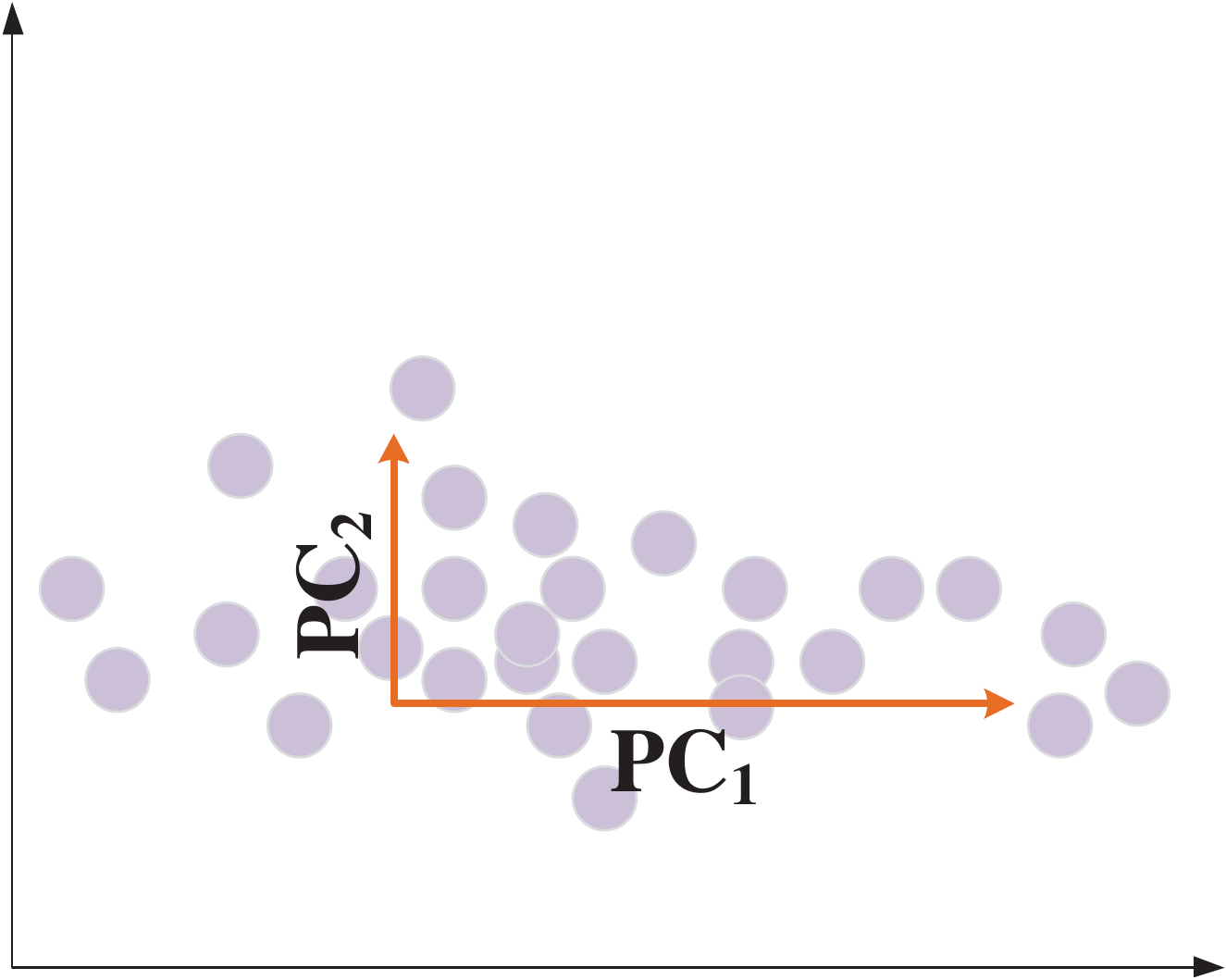}
  }
  \caption{Using \PCA to reduce dimensionality of a three-dimensional feature space. The principal components are firstly computed
  (a). Then the first two principal components ($PC_1$ and $PC_2$) are selected to represent the original three-dimensional feature space on a new two-dimensional
  space {b}.  \label{fig:pca_example}}
  \vspace{-3mm}
\end{figure}

Figure~\ref{fig:pca_example} demonstrates the use of \PCA to reduce the number of dimensions. The input in this example is a
three-dimensional space defined by $M_1$, $M_2$ and $M_3$, as shown in Figure~\ref{fig:pca_example} (a). Three components: $PC_1$, $PC_2$
and $PC_3$, which account for the variance of the data, are firstly calculated. Here, $PC_1$ and $PC_2$ contribute most to the variance of
the data and $PC_3$ accounts for the least variance. Using only $PC_1$ and $PC_2$, one can transform the original, three-dimensional space
into a new, two-dimensional coordinate system (as illustrated in Figure~\ref{fig:pca_example}b) while preserving much of the variance of
the original data.

\PCA has been used in many prior compiler research works for feature reduction~\cite{
1106006,Agakov:2006:UML:1121992.1122412,Dubach:2007:FCO:1242531.1242553,Chen:2010:EIO:1806596.1806647,Wang:2010:PSP:1854273.1854313,
ashouri2014bayesian,thomson2009reducing,Magni:2014:AOT:2628071.2628087}. It has also been used in prior works  to visualise the working
mechanism of a machine learning model, e.g. to show how benchmarks can be grouped in the feature space~\cite{middleware17}, by projecting
features from a high-dimensional space into a 2-dimensional space.

We want to stress that \PCA does not select some features and discard the others. Instead, it linearly combines the original features to
construct new features that can summarise the list of the original features. \PCA is useful when there is some redundancy in the raw
features, i.e. some of the features are correlated with one another. Similar feature reduction methods include factor analysis and linear
discriminant analysis (LDA), which all try to reduce the number of features by linearly combining multiple raw features. However, \PCA
seems to be the most popular feature reduction method used in
 compiler research, probably due to its simplicity.

An alternative way of reducing the number of features used is via an  \emph{autoencoder}~\cite{Bengio:2009:LDA:1658423.1658424}. It  is a
neural network that finds a representation (encoding) for a set of data, by dimensionality reduction. Autoencoders works by learning an
encoder and a decoder from the input data. The encoder tries to compress the original input into a low-dimensional representation, while
the decoder tries to reconstruct the original input based on the low-dimension representations generated by the encoder. As a result, the
autoencoder has been widely used to remove the data noise as well to reduce the data dimension~\cite{deng2010binary}.

Autoencoders have been applied to various natural language processing tasks~\cite{Vincent:2008:ECR:1390156.1390294}, often being used
together with \DNNs. Recently, it has been employed to model program source code to obtain a compact set of features that can characterise
the input program source~\cite{Mou2013, White2016,Cummins:2017:SBP:3049832.3049843,pact17, White2017}.


%% file: scope.tex
\section{Scope}
Machine learning has been used to solve a wide range of problems, from the early successful work of selecting compiler flags for sequential
programs, to recent works on scheduling and optimising parallel programs on heterogeneous multi-cores. In this section, we review the types
of problems that have been exploited in prior works.

\subsection{Optimise sequential programs}
Early works for machine learning in compilers look at how, or if, a compiler optimisation should be applied to a sequential program. Some
of the previous studies build supervised classifiers to predict the optimal loop unroll
factor~\cite{Leather:2009:AFG:1545006.1545059,Stephenson:2005:PUF:1048922.1048981}  or to determine whether a function should be
inlined~\cite{Simon:2013:ACI:2495258.2495914,Cavazos:2005:ATI:1105760.1105779}. These works target a fixed set of compiler options, by
representing the optimisation problem as a multi-class classification problem -- where each compiler option is a class. For example,
Leather \etal~\cite{ Leather:2009:AFG:1545006.1545059} considered a loop unroll factor between 0 and 15 (16 configurations in total),
treating each candidate unroll factor as a class; they compiled and profiled each training program by trying all 16 configurations to find
out the best loop unroll factor for each program, and then learned a decision tree model from the training data.

There are other compiler problems where the number of possible options is massive. For instance, the work presented in
\cite{Dubach:2007:FCO:1242531.1242553} considers 54 code transformations of GCC. While these options are only a subset from the over 100s
transformations provided by GCC, the resulted combinatorial compiler configurations lead to a space of approximately $10^{34}$. Although it
is possible to build a classifier to directly predict the optimal setting from a large space, to learn an effective model would require a
large volume of training programs in order to have an adequate sampling over the space. Doing so is difficult because (a) there are only a
 few dozen common benchmarks available and (b) compiler developers need to generate the training data themselves.

Evolutionary algorithms like generic search are often used to explore a large design space (see also Section~\ref{sec:ea}). Prior works
have used evolutionary algorithms to solve the phase ordering problem (i.e. at which order a set of compiler transformations should be
applied)~\cite{Almagor:2004:FEC:997163.997196,Cooper:2005:AAC:1065910.1065921,Ashouri:2017:MMC:3132652.3124452}, determining the compiler
flags during iterative compilation~\cite{cooper2002adaptive,5688317,Fursin:2010:COP:1880043.1880047,Kukunas:2010:GAI:1830761.1830879},
selecting loop transformations~\cite{Pouchet:2008:IOP:1375581.1375594}, and tuning algorithmic
choices~\cite{ansel:pldi:2009,ansel:cgo:2011}, etc.

\subsection{Optimise parallel programs \label{sec:parallel_opt}}
How to effectively optimise parallel programs has received significant attentions in the past decade, largely because the hardware industry
has adopted multi-core design to avoid the power wall~\cite{asanovic2006landscape}. While multi- and many-core architectures provide the
potential for high performance and energy-efficient computing, the potential performance can only be unlocked if the application programs
are suitably parallel and can be made to match the underlying heterogeneous platform. Without this, the myriad cores on multi-core
processors and their specialised processing elements will sit idle or poorly utilised. To this end, researchers have extended the reach of
machine learning to optimise parallel programs.

A line of research in parallel program optimisation is parallelism mapping. That is, given an already parallelised program, how to map the
application parallelism to match the underlying hardware to make the program runs as fast as possible or be as energy-efficient as
possible. Zhang \etal developed a decision tree based approach to predict the scheduling policy to use for an OpenMP parallel
region~\cite{1419864}. The work presented in \cite{Wang:2009:MPM:1504176.1504189} employs two machine learning techniques to predict the
optimal number of threads as well as the scheduling policy to use for OpenMP parallel loop. Specifically, it uses a regression-based \ANN
model to predict the speedup of a parallel loop when it runs with a given number of threads (to search for the optimal number threads), and
a \SVM classifer to predict the scheduling policy. There are also works use machine learning to determine the optimum degree of parallelism
for transactional memory~\cite{6298188} and hardware source allocation~\cite{Delimitrou:2014:QRQ:2541940.2541941}, or to select a code
version from a pool of choices to use~\cite{5395311}. Castro \etal developed a decision tree classifier to predict the thread mapping
strategy in the context of software transactional memory~\cite{6152736}. Jung \etal constructed a \ANN based predictor to select an
effective data structure on a specific micro-architecture~\cite{ Jung:2011:BES:1993498.1993509}.

The work presented in \cite{Wang:2010:PSP:1854273.1854313} and \cite{Wang:2013:UML:2509420.2512436} is a unique approach for applying
machine learning to map complex parallel programs with unbounded parallel graph structures.  The work considers the question of finding the
optimal graph structure of a streaming program.  The idea was that rather than trying to predict a sequence of transformations over an
unbounded graph, where legality and consistency is a real problem, we should consider the problem from the dual feature space. The work
showed that it is possible to predict the best target feature (i.e. the characteristics that an ideal transformed program should have)
which then can be used to evaluate the worth of candidate transformed graphs (without compiling and profiling the resulted graphs) in the
original feature space.

The Petabricks project~\cite{ansel:pldi:2009, chan:sc:2009, pacula:evoapps:2012}
 takes an evolutionary approach for program tuning. The Petabricks compiler
employs genetic search algorithms to tune algorithmic choices. Due to the expensive overhead of the search, much of auto-tuning is done at
static compile time. Their work shows that one can utilise the idle processors on a multi-core systems to perform online tuning~\cite{
Ansel:2012:SOA:2380403.2380425}, where half of the cores are devoted to a known safe program configuration, while the other half are used
for an experimental program configuration. In this way, when the results of the faster configuration are returned, the slower version will
be terminated.

The idea of combining compile-time knowledge and runtime information to achieve better optimizations has been exploited by the ADAPT
compiler~\cite{Voss:2001:HAP:379539.379583}. Using the ADAPT compiler, users describe what optimizations are available and provide
heuristics for applying these optimizations. The compiler then reads these descriptions and generates application-specific runtime systems
to apply the heuristics. Runtime code tuning is also exploited by Active Harmony~\cite{tiwari2011online}, which utilises the computing
resources in HPC systems to evaluate different code-variants on different nodes to find the best-performing version.

There is also an extensive body of work on how to optimise programs on heterogeneous multi-core systems. One of the problems for
heterogeneous multi-core optimisation is to determine when and how to use the heterogeneous processors. Researchers have used machine
learning to build classifiers to determine which processor to use~\cite{6494993} and at which clock frequency the processor should
operate~\cite{Taylor2017AdaptiveOF,infocom17}. Others used regression techniques to build curve fitting models to search for the sweat spot
for work partitioning among processors~\cite{Luk:2009:QEP:1669112.1669121} or a trade-off of energy and performance~\cite{zhu2013high}.

Another line of research combines compiler-based analysis and machine learning to optimise programs in in the presence of competing
workloads. This research problem is important because programs rarely run in isolation and must share the computing resources with other
co-running workloads. In \cite{Wang:2013:SAM:2495258.2495918} and \cite{Grewe:2011:WMA:1944862.1944881}, an \ANN model based on static code
features and runtime information was built to predict the number of threads to use for a target program when it runs with external
workloads. Later in \cite{Emani:2015:CDM:2737924.2737999} an ensemble learning based approach was used, which leads to significantly better
performance over \cite{Wang:2013:SAM:2495258.2495918}. In \cite{Emani:2015:CDM:2737924.2737999} several models are firstly trained offline;
and then one of the model is selected at runtime, taking into consideration the competing workloads and available hardware resources. The
central idea is that instead of using a single monolithic model, we can use multiple models where each model is specialised for mode ling a
subset of applications or a particular runtime scenario. Using this approach, a model is used when its predictions are effective.

Some recent works developed machine learning models based on static code features and dynamic runtime information to schedule OpenCL
programs in the presence of GPU contention. The work presented in ~\cite{grewe2013opencl} uses \SVM classification to predict the work
partition ratio between the CPU and GPU when multiple programs are competing to run on a single GPU. The work described in \cite{yw14} aims
to improve the overall system throughput when there are multiple OpenCL programs competing to run on the GPU. They developed an \ANN model
to predict the potential speedup for running an OpenCL kernel on the GPU. The speedup prediction is then used as a proxy to determine which
of the waiting OpenCL tasks get to run on the GPU and at which order.

The approaches presented in \cite{Tang:2012:CNM:2259016.2259018} and \cite{ Tang:2013:RRS:2451116.2451126} target task co-location in a
data centre environment.  They use compiler based code transformations to reduce the contention for multiple co-running tasks. A linear
regression model was employed to calculate the contention score of code regions based on performance counter values. Then, a set of
compiler-based code transformations is applied to reduce the resource demands of highly contentious code.

\subsection{Other research problems \label{sec:others}}
There are many works have demonstrated that machine learning is a powerful technique in performance and cost modelling
~\cite{Matsunaga:2010:UML:1844765.1845166, venkataraman2016ernest, Sankaran:2016:PMB:2903150.2911714, Kang:2017:NCI:3037697.3037698}, and
in task and resource scheduling~\cite{Zhang:2014:SPQ:2742155.2742197,Delimitrou:2014:QRQ:2541940.2541941, petrucci2015octopus,
yadwadkar2016multi}. We envision that many of these techniques can be used to provide evidences to support runtime program optimizations
through e.g. just-in-time compilation.

While not directly target code optimisation, compiler based code analysis and machine learning techniques have been used in conjunction to
solve various software engineering tasks. These include detecting code similarities~\cite{David:2014:TCS:2594291.2594343,
David:2016:SSB:2908080.2908126}, automatic comment generation~\cite{wong2015clocom}, mining API usage
patterns~\cite{Fowkes:2016:PPA:2950290.2950319, Nguyen:2016:ACR:2950290.2950333}, predicting program
properties~\cite{Raychev:2016:PMC:2983990.2984041}, code de-obfuscation for malware detection~\cite{ Bichsel:2016:SDA:2976749.2978422},
etc. It is worth mentioning that many of these recent works show that the past development knowledge extracted from large code bases such
as GitHub are valuable for learning an effective model.  There were two recent studies performed by Cummins \etal, which mine Github to
synthesize OpenCL benchmarks~\cite{Cummins:2017:SBP:3049832.3049843} and code extract features from source code~\cite{pact17}. Both studies
demonstrate the usefulness of large code bases and deep learning techniques for learning predictive models for compiler optimizations. We
envision that the rich information in large open source code bases could provide a powerful knowledge base for training machine learning
models to solve compiler optimisation problems, and deep learning could be used as an effective tool to extract such knowledge from massive
program source code.

%% file: challenges.tex
\section{Discussion \label{sec:discussion}}
One of the real benefits of machine learning based approaches is that it forces an empirical driven  approach to compiler construction. New
models have to be based on empirical data which can then be verified by independent experimentation. This experiment -- hypothesis -- test
cycle is well known in the physical sciences but is a relatively new addition compiler construction.

As machine learning based techniques require a sampling of the optimisation space for training data, we typically know the best
optimisation for any program in the training set. If we exclude this benchmark from training, we therefore have access to an upper bound on
performance or oracle for this program.  This immediately lets us know how good existing techniques are. If they are 50\% of this optimum
or 95\% of this optimum immediately tells us whether the problem  is worth exploring.

Furthermore we can construct naive techniques -- e.g. a random optimisation and see its performance. If this performed a number of times,
it will have an expected value of the mean of the optimisation speedups. We can then demand that any new heuristic should outperform this
-- though in our experience there have been cases where state-of the art work was actually less than random.

\subsection{Not a panacea}

This article has by and large been very upbeat about the use of
machine learning.
However, there are number of hurdles to overcome to make it a practical
reality and opens up new questions about optimisation

Training cost is an issue that many find alarming. In practise the cost is much less than a compiler writer and techniques like active
learning can be employed to reduce overhead of training data
generation~\cite{balaprakash2013active,ogilvie2014fast,zuluaga2013active,Ogilvie:2017:MCI:3049832.3049859}. Although its true to say that
generating many differently compiled programs executing and timing them is entirely automatic, finding the right data requires careful
consideration.  If the optimizations explored have little positive performance on the programs then there is nothing worth learning.

The most immediate problem continues to be  gathering enough sufficient high quality training data. Although there are numerous benchmark
sites publicly available, the number of programs available is relatively sparse compared to to the number a typical compiler will encounter
in its lifetime. This is particular true in specialist domains where there may not be any public benchmarks. Automatic benchmark generation
work will help here, but existing approaches do not guarantee that the generated benchmarks effectively represent the design space.
Therefore, the larger issue of the structure of the the program space remains.

A really fundamental problem is that if  we build our optimisation models based purely on empirical data, then we must guarantee that this
data is correct and representative; we must learn the signal not the noise. Peer review of machine learning  approach is difficult. Black
box modelling prevents the quality of the model from being questioned unlike handcrafted heuristics. In a sense reviewers now have to
scrutinise that the experiments were fairly done.  This  means all training and test data must be publicly available for scrutiny. This is
common practise in other empirical sciences. The artefact evaluation committee is an example of this~\cite{AEC,AEC2}.

Although the ability to automatically learn how to best optimise an application and adapt to change is a big step forward, machine learning
can only learn form what is provided by the compiler writer. Machine learning can not invent new program transformations to apply nor can
it derive analysis that determines whether or not a transformation is legal -- all of this is beyond its scope.

\subsection{Will this put compiler writers out of a job?}

In fact machine learning based compilation will paradoxically lead to a renaissance in compiler optimisation. Compiler have become so
complex that adding a new optimisation or compiler phase can lead to performance regressions. This in turn has led to a conservative mind
set where new transformations are not considered if they may rock the boat.  The core issue is that systems are so complex that is
impossible to know for sure when to use or not such an optimisation. Machine learning can remove this uncertainty by automatically
determining when an optimisation is profitable. This now frees the compiler writer to develop ever more sophisticated techniques. He/she
does not need to worry about how they interfere with other optimizations -- machine learning  looks after this. We can now develop
optimizations that will typically only work for specific domains, and not worry about coordinating their integration into a general purpose
system. It allows different communities to develop novel optimizations and naturally integrate them. So rather than closing down the
opportunity for new ideas, it opens up new vistas.

\subsection{Open research directions}

Machine learning has demonstrated its utility as a means of automating compiler profitability analysis. It will continue to be used for
more complex optimisation problems and is likely to be the default approach to selecting compiler optimizations in the coming decade.

The open research directions go beyond predicting the best optimizations to apply. One central issue is what does the program space look
like? We know that programs with linear array accesses inside perfect loop nests need different treatment compared to, say, distributed
graph processing programs. If we could have a  map that allows us to measure distances between programs, then we could see whether there
are regions that are well served by compiler characterise and  other regions that are sparse and currently ignored. If we could do the same
for hardware, then we may be better able to design hardware likely to be of use for emerging applications.

Can machine learning also be applied to compiler analysis? For instance is it possible to learn dataflow or point-to analysis?  As deep
learning has the ability to automatically constructs features, can we find a set of features that are common across all optimizations and
analyses. Can we learn the ideal compiler intermediate representation? There is a wide range of interesting research questions that remains
unexplored.

%% file: conclusions.tex
\section{Conclusion \label{sec:conclusion}}
This paper has introduced machine learning based compilation and described its power in determining an  evidence based approach to compiler
optimisation. It is the latest stage in   fifty years of compiler automation. Machine learning based compilation is now a   mainstream
compiler research area and over the last decade or so, has generated a large amount of academic interest and papers. While it is impossible
to provide a definitive cataloguer  of all research, we have tried to provide a comprehensive and accessible survey of the main research
areas and future directions. Machine learning is not a panacea. It can only learn the data we provide. Rather than, as some fear, it dumbs
down the role of compiler writers, it opens up the possibility of much greater creativity and new research areas.

%% file: main.bbl
\begin{thebibliography}{100}
\providecommand{\url}[1]{#1}
\csname url@samestyle\endcsname
\providecommand{\newblock}{\relax}
\providecommand{\bibinfo}[2]{#2}
\providecommand{\BIBentrySTDinterwordspacing}{\spaceskip=0pt\relax}
\providecommand{\BIBentryALTinterwordstretchfactor}{4}
\providecommand{\BIBentryALTinterwordspacing}{\spaceskip=\fontdimen2\font plus
\BIBentryALTinterwordstretchfactor\fontdimen3\font minus
  \fontdimen4\font\relax}
\providecommand{\BIBforeignlanguage}[2]{{%
\expandafter\ifx\csname l@#1\endcsname\relax
\typeout{** WARNING: IEEEtran.bst: No hyphenation pattern has been}%
\typeout{** loaded for the language `#1'. Using the pattern for}%
\typeout{** the default language instead.}%
\else
\language=\csname l@#1\endcsname
\fi
#2}}
\providecommand{\BIBdecl}{\relax}
\BIBdecl

\bibitem{chipps1956mathematical}
J.~Chipps, M.~Koschmann, S.~Orgel, A.~Perlis, and J.~Smith, ``A mathematical
  language compiler,'' in \emph{Proceedings of the 1956 11th ACM national
  meeting}.\hskip 1em plus 0.5em minus 0.4em\relax ACM, 1956, pp. 114--117.

\bibitem{sheridan1959arithmetic}
P.~B. Sheridan, ``The arithmetic translator-compiler of the ibm fortran
  automatic coding system,'' \emph{Communications of the ACM}, vol.~2, no.~2,
  pp. 9--21, 1959.

\bibitem{mcilroy1960macro}
M.~D. McIlroy, ``Macro instruction extensions of compiler languages,''
  \emph{Communications of the ACM}, vol.~3, no.~4, pp. 214--220, 1960.

\bibitem{gauci2010machine}
A.~Gauci, K.~Z. Adami, and J.~Abela, ``Machine learning for galaxy morphology
  classification,'' \emph{arXiv preprint arXiv:1005.0390}, 2010.

\bibitem{schoen2013power}
H.~Schoen, D.~Gayo-Avello, P.~Takis~Metaxas, E.~Mustafaraj, M.~Strohmaier, and
  P.~Gloor, ``The power of prediction with social media,'' \emph{Internet
  Research}, vol.~23, no.~5, pp. 528--543, 2013.

\bibitem{milepostgcc:slashdot}
\BIBentryALTinterwordspacing
Slashdot. (2009) {IBM} releases open source machine learning compiler.
  [Online]. Available:
  \url{https://tech.slashdot.org/story/09/07/03/0143233/ibm-releases-open-source-machine-learning-compiler}
\BIBentrySTDinterwordspacing

\bibitem{massalin1987superoptimizer}
H.~Massalin, ``Superoptimizer: a look at the smallest program,'' in \emph{ACM
  SIGPLAN Notices}, vol.~22, no.~10, 1987, pp. 122--126.

\bibitem{ivory1825method}
J.~Ivory, ``I. on the method of the least squares,'' \emph{The Philosophical
  Magazine and Journal: Comprehending the Various Branches of Science, the
  Liberal and Fine Arts, Agriculture, Manufactures and Commerce}, vol.~65, no.
  321, pp. 3--10, 1825.

\bibitem{adcock1878problem}
R.~J. Adcock, ``A problem in least squares,'' \emph{The Analyst}, vol.~5,
  no.~2, pp. 53--54, 1878.

\bibitem{datta2008stencil}
K.~Datta, M.~Murphy, V.~Volkov, S.~Williams, J.~Carter, L.~Oliker,
  D.~Patterson, J.~Shalf, and K.~Yelick, ``Stencil computation optimization and
  auto-tuning on state-of-the-art multicore architectures,'' in
  \emph{Proceedings of the 2008 ACM/IEEE conference on Supercomputing}, 2008,
  p.~4.

\bibitem{ansel:cgo:2011}
J.~Ansel, Y.~L.~W. ans Cy~Chan, M.~Olszewski, A.~Edelman, and S.~Amarasinghe,
  ``Language and compiler support for auto-tuning variable-accuracy
  algorithms,'' in \emph{The International Symposium on Code Generation and
  Optimization}, ser. CGO '11, 2011.

\bibitem{kurzak2016implementation}
J.~Kurzak, H.~Anzt, M.~Gates, and J.~Dongarra, ``Implementation and tuning of
  batched cholesky factorization and solve for nvidia gpus,'' \emph{IEEE
  Transactions on Parallel and Distributed Systems}, vol.~27, no.~7, 2016.

\bibitem{tsai2016performance}
Y.~M. Tsai, P.~Luszczek, J.~Kurzak, and J.~Dongarra, ``Performance-portable
  autotuning of opencl kernels for convolutional layers of deep neural
  networks,'' in \emph{Workshop on Machine Learning in HPC Environments
  (MLHPC)}, 2016, pp. 9--18.

\bibitem{lesk1975lex}
M.~E. Lesk and E.~Schmidt, ``Lex: A lexical analyzer generator,'' 1975.

\bibitem{johnson1975yacc}
S.~C. Johnson, \emph{Yacc: Yet another compiler-compiler}.\hskip 1em plus 0.5em
  minus 0.4em\relax Bell Laboratories Murray Hill, NJ, 1975, vol.~32.

\bibitem{monsifrot2002machine}
A.~Monsifrot, F.~Bodin, and R.~Quiniou, ``A machine learning approach to
  automatic production of compiler heuristics,'' in \emph{International
  Conference on Artificial Intelligence: Methodology, Systems, and
  Applications}, 2002, pp. 41--50.

\bibitem{Magni:2014:AOT:2628071.2628087}
A.~Magni, C.~Dubach, and M.~O'Boyle, ``Automatic optimization of
  thread-coarsening for graphics processors,'' in \emph{Proceedings of the 23rd
  International Conference on Parallel Architectures and Compilation}, ser.
  PACT '14, 2014, pp. 455--466.

\bibitem{Unkule:2012:ARG:2259230.2259233}
S.~Unkule, C.~Shaltz, and A.~Qasem, ``Automatic restructuring of {GPU} kernels
  for exploiting inter-thread data locality,'' in \emph{Proceedings of the 21st
  International Conference on Compiler Construction}, ser. CC'12, 2012, pp.
  21--40.

\bibitem{Volkov:2008:BGT:1413370.1413402}
V.~Volkov and J.~W. Demmel, ``Benchmarking {GPUs} to tune dense linear
  algebra,'' in \emph{Proceedings of the 2008 ACM/IEEE Conference on
  Supercomputing}, ser. SC '08, 2008, pp. 31:1--31:11.

\bibitem{Yang:2012:UOC:2207222.2207225}
Y.~Yang, P.~Xiang, J.~Kong, M.~Mantor, and H.~Zhou, ``A unified optimizing
  compiler framework for different gpgpu architectures,'' \emph{ACM Trans.
  Archit. Code Optim.}, vol.~9, no.~2, pp. 9:1--9:33, 2012.

\bibitem{Lattner:2004:LCF:977395.977673}
C.~Lattner and V.~Adve, ``{LLVM}: A compilation framework for lifelong program
  analysis \& transformation,'' in \emph{Proceedings of the International
  Symposium on Code Generation and Optimization: Feedback-directed and Runtime
  Optimization}, ser. CGO '04, 2004.

\bibitem{bodin1998iterative}
F.~Bodin, T.~Kisuki, P.~Knijnenburg, M.~O'Boyle, and E.~Rohou, ``Iterative
  compilation in a non-linear optimisation space,'' in \emph{Workshop on
  Profile and Feedback-Directed Compilation}, 1998.

\bibitem{knijnenburg2003combined}
P.~M. Knijnenburg, T.~Kisuki, and M.~F. O'Boyle, ``Combined selection of tile
  sizes and unroll factors using iterative compilation,'' \emph{The Journal of
  Supercomputing}, vol.~24, no.~1, pp. 43--67, 2003.

\bibitem{FFTW05}
M.~Frigo and S.~G. Johnson, ``The design and implementation of {FFTW3},''
  \emph{Proceedings of the IEEE}, vol.~93, no.~2, pp. 216--231, 2005, special
  issue on ``Program Generation, Optimization, and Platform Adaptation''.

\bibitem{Agakov:2006:UML:1121992.1122412}
F.~Agakov, E.~Bonilla, J.~Cavazos, B.~Franke, G.~Fursin, M.~F.~P. O'Boyle,
  J.~Thomson, M.~Toussaint, and C.~K.~I. Williams, ``Using machine learning to
  focus iterative optimization,'' in \emph{Proceedings of the International
  Symposium on Code Generation and Optimization}, ser. CGO '06, 2006, pp.
  295--305.

\bibitem{Nobre:2016:GIC:2907950.2907959}
R.~Nobre, L.~G.~A. Martins, and J.~a. M.~P. Cardoso, ``A graph-based iterative
  compiler pass selection and phase ordering approach,'' in \emph{Proceedings
  of the 17th ACM SIGPLAN/SIGBED Conference on Languages, Compilers, Tools, and
  Theory for Embedded Systems}, ser. LCTES 2016, 2016, pp. 21--30.

\bibitem{leupers1999function}
R.~Leupers and P.~Marwedel, ``Function inlining under code size constraints for
  embedded processors,'' in \emph{Computer-Aided Design, 1999. Digest of
  Technical Papers. 1999 IEEE/ACM International Conference on}.\hskip 1em plus
  0.5em minus 0.4em\relax IEEE, 1999, pp. 253--256.

\bibitem{Cooper:2008:ASI:1788374.1788381}
K.~D. Cooper, T.~J. Harvey, and T.~Waterman, ``An adaptive strategy for inline
  substitution,'' in \emph{Proceedings of the Joint European Conferences on
  Theory and Practice of Software 17th International Conference on Compiler
  Construction}, ser. CC'08/ETAPS'08, 2008, pp. 69--84.

\bibitem{Simon:2013:ACI:2495258.2495914}
D.~Simon, J.~Cavazos, C.~Wimmer, and S.~Kulkarni, ``Automatic construction of
  inlining heuristics using machine learning,'' in \emph{Proceedings of the
  2013 IEEE/ACM International Symposium on Code Generation and Optimization
  (CGO)}, ser. CGO '13, 2013, pp. 1--12.

\bibitem{zhao2004inline}
P.~Zhao and J.~Amaral, ``To inline or not to inline? enhanced inlining
  decisions,'' \emph{Languages and Compilers for Parallel Computing}, pp.
  405--419, 2004.

\bibitem{Wagner:1994:ASE:178243.178251}
T.~A. Wagner, V.~Maverick, S.~L. Graham, and M.~A. Harrison, ``Accurate static
  estimators for program optimization,'' in \emph{Proceedings of the ACM
  SIGPLAN 1994 Conference on Programming Language Design and Implementation},
  ser. PLDI '94, 1994, pp. 85--96.

\bibitem{629825}
V.~Tiwari, S.~Malik, and A.~Wolfe, ``Power analysis of embedded software: A
  first step towards software power minimization,'' in \emph{IEEE/ACM
  International Conference on Computer-Aided Design}, 1994, pp. 384--390.

\bibitem{Cooper1999}
K.~D. Cooper, P.~J. Schielke, and D.~Subramanian, ``Optimizing for reduced code
  space using genetic algorithms,'' in \emph{Proceedings of the ACM SIGPLAN
  1999 Workshop on Languages, Compilers, and Tools for Embedded Systems}, ser.
  LCTES '99, 1999, pp. 1--9.

\bibitem{metaopt}
M.~Stephenson, S.~Amarasinghe, M.~Martin, and U.-M. O'Reilly, ``Meta
  optimization: Improving compiler heuristics with machine learning,'' in
  \emph{Proceedings of the ACM SIGPLAN 2003 Conference on Programming Language
  Design and Implementation}, ser. PLDI '03, 2003, pp. 77--90.

\bibitem{Cavazos:2005:ATI:1105760.1105779}
J.~Cavazos and M.~F.~P. O'Boyle, ``Automatic tuning of inlining heuristics,''
  in \emph{Proceedings of the 2005 ACM/IEEE Conference on Supercomputing}, ser.
  SC '05, 2005.

\bibitem{Hoste:2008:CCO:1356058.1356080}
K.~Hoste and L.~Eeckhout, ``Cole: Compiler optimization level exploration,'' in
  \emph{Proceedings of the 6th Annual IEEE/ACM International Symposium on Code
  Generation and Optimization}, ser. CGO '08, 2008, pp. 165--174.

\bibitem{Kim2004}
M.~Kim, T.~Hiroyasu, M.~Miki, and S.~Watanabe, \emph{SPEA2+: Improving the
  Performance of the Strength Pareto Evolutionary Algorithm 2}, 2004, pp.
  742--751.

\bibitem{Luk:2009:QEP:1669112.1669121}
C.-K. Luk, S.~Hong, and H.~Kim, ``Qilin: Exploiting parallelism on
  heterogeneous multiprocessors with adaptive mapping,'' in \emph{Proceedings
  of the 42Nd Annual IEEE/ACM International Symposium on Microarchitecture},
  ser. MICRO 42, 2009, pp. 45--55.

\bibitem{yw14}
Y.~Wen, Z.~Wang, and M.~O'Boyle, ``Smart multi-task scheduling for {OpenCL}
  programs on {CPU/GPU} heterogeneous platforms,'' in \emph{21st Annual IEEE
  International Conference on High Performance Computing (HiPC 2014)}.\hskip
  1em plus 0.5em minus 0.4em\relax IEEE, 2014.

\bibitem{Brewer:1995:HOV:209936.209946}
E.~A. Brewer, ``High-level optimization via automated statistical modeling,''
  in \emph{Proceedings of the Fifth ACM SIGPLAN Symposium on Principles and
  Practice of Parallel Programming}, ser. PPOPP '95, 1995, pp. 80--91.

\bibitem{4145110}
K.~Vaswani, M.~J. Thazhuthaveetil, Y.~N. Srikant, and P.~J. Joseph,
  ``Microarchitecture sensitive empirical models for compiler optimizations,''
  in \emph{International Symposium on Code Generation and Optimization
  (CGO'07)}, 2007, pp. 131--143.

\bibitem{Lee:2006:AER:1168857.1168881}
B.~C. Lee and D.~M. Brooks, ``Accurate and efficient regression modeling for
  microarchitectural performance and power prediction,'' in \emph{Proceedings
  of the 12th International Conference on Architectural Support for Programming
  Languages and Operating Systems}, ser. ASPLOS XII, 2006, pp. 185--194.

\bibitem{Park:2011:PMP:2190025.2190059}
E.~Park, L.-N. Pouche, J.~Cavazos, A.~Cohen, and P.~Sadayappan, ``Predictive
  modeling in a polyhedral optimization space,'' in \emph{Proceedings of the
  9th Annual IEEE/ACM International Symposium on Code Generation and
  Optimization}, ser. CGO '11, 2011, pp. 119--129.

\bibitem{Curtis-Maury:2008:PMM:1454115.1454151}
M.~Curtis-Maury, A.~Shah, F.~Blagojevic, D.~S. Nikolopoulos, B.~R. de~Supinski,
  and M.~Schulz, ``Prediction models for multi-dimensional power-performance
  optimization on many cores,'' in \emph{Proceedings of the 17th International
  Conference on Parallel Architectures and Compilation Techniques}, ser. PACT
  '08, 2008, pp. 250--259.

\bibitem{Singh2010}
K.~Singh, M.~Curtis-Maury, S.~A. McKee, F.~Blagojevi{\'{c}}, D.~S.
  Nikolopoulos, B.~R. de~Supinski, and M.~Schulz, \emph{Comparing Scalability
  Prediction Strategies on an SMP of CMPs}, 2010, pp. 143--155.

\bibitem{Wang:2009:MPM:1504176.1504189}
Z.~Wang and M.~F. O'Boyle, ``Mapping parallelism to multi-cores: A machine
  learning based approach,'' in \emph{Proceedings of the 14th ACM SIGPLAN
  Symposium on Principles and Practice of Parallel Programming}, ser. PPoPP
  '09, 2009, pp. 75--84.

\bibitem{Kang:2017:NCI:3037697.3037698}
Y.~Kang, J.~Hauswald, C.~Gao, A.~Rovinski, T.~Mudge, J.~Mars, and L.~Tang,
  ``Neurosurgeon: Collaborative intelligence between the cloud and mobile
  edge,'' in \emph{Proceedings of the Twenty-Second International Conference on
  Architectural Support for Programming Languages and Operating Systems}, ser.
  ASPLOS '17, 2017, pp. 615--629.

\bibitem{Benini:1998:RMB:1275815.1275817}
L.~Benini, A.~Bogliolo, M.~Favalli, and G.~De~Micheli, ``Regression models for
  behavioral power estimation,'' \emph{Integr. Comput.-Aided Eng.}, vol.~5,
  no.~2, pp. 95--106.

\bibitem{6089692}
S.~K. Rethinagiri, R.~B. Atitallah, and J.~L. Dekeyser, ``A system level power
  consumption estimation for mpsoc,'' in \emph{2011 International Symposium on
  System on Chip (SoC)}, 2011, pp. 56--61.

\bibitem{Schurmans:2016:FEP:3008024.2987375}
S.~Sch\"{u}rmans, G.~Onnebrink, R.~Leupers, G.~Ascheid, and X.~Chen,
  ``Frequency-aware esl power estimation for arm cortex-a9 using a black box
  processor model,'' \emph{ACM Trans. Embed. Comput. Syst.}, vol.~16, no.~1,
  pp. 26:1--26:26, 2016.

\bibitem{4629274}
M.~Curtis-Maury, K.~Singh, S.~A. McKee, F.~Blagojevic, D.~S. Nikolopoulos,
  B.~R. de~Supinski, and M.~Schulz, ``Identifying energy-efficient concurrency
  levels using machine learning,'' in \emph{2007 IEEE International Conference
  on Cluster Computing}, 2007, pp. 488--495.

\bibitem{Stephenson:2005:PUF:1048922.1048981}
M.~Stephenson and S.~Amarasinghe, ``Predicting unroll factors using supervised
  classification,'' in \emph{Proceedings of the International Symposium on Code
  Generation and Optimization}, ser. CGO '05, 2005, pp. 123--134.

\bibitem{Cavazos:2007:RSG:1251974.1252540}
J.~Cavazos, G.~Fursin, F.~Agakov, E.~Bonilla, M.~F.~P. O'Boyle, and O.~Temam,
  ``Rapidly selecting good compiler optimizations using performance counters,''
  in \emph{Proceedings of the International Symposium on Code Generation and
  Optimization}, ser. CGO '07, 2007.

\bibitem{Cavazos:2006:MDC:1167473.1167492}
J.~Cavazos and M.~F.~P. O'Boyle, ``Method-specific dynamic compilation using
  logistic regression,'' in \emph{Proceedings of the 21st Annual ACM SIGPLAN
  Conference on Object-oriented Programming Systems, Languages, and
  Applications}, ser. OOPSLA '06, 2006, pp. 229--240.

\bibitem{Dubach:2007:FCO:1242531.1242553}
C.~Dubach, J.~Cavazos, B.~Franke, G.~Fursin, M.~F. O'Boyle, and O.~Temam,
  ``Fast compiler optimisation evaluation using code-feature based performance
  prediction,'' in \emph{Proceedings of the 4th International Conference on
  Computing Frontiers}, ser. CF '07, 2007, pp. 131--142.

\bibitem{Yuki:2010:ACT:1772954.1772982}
T.~Yuki, L.~Renganarayanan, S.~Rajopadhye, C.~Anderson, A.~E. Eichenberger, and
  K.~O'Brien, ``Automatic creation of tile size selection models,'' in
  \emph{Proceedings of the 8th Annual IEEE/ACM International Symposium on Code
  Generation and Optimization}, ser. CGO '10, 2010, pp. 190--199.

\bibitem{6406709}
A.~M. Malik, ``Optimal tile size selection problem using machine learning,'' in
  \emph{2012 11th International Conference on Machine Learning and
  Applications}, vol.~2, 2012, pp. 275--280.

\bibitem{Moore2014}
R.~W. Moore and B.~R. Childers, ``Building and using application utility models
  to dynamically choose thread counts,'' \emph{The Journal of Supercomputing},
  vol.~68, no.~3, pp. 1184--1213, 2014.

\bibitem{5160988}
Y.~Liu, E.~Z. Zhang, and X.~Shen, ``A cross-input adaptive framework for {GPU}
  program optimizations,'' in \emph{2009 IEEE International Symposium on
  Parallel Distributed Processing}, 2009, pp. 1--10.

\bibitem{Perelman:2003:USA:781027.781076}
E.~Perelman, G.~Hamerly, M.~Van~Biesbrouck, T.~Sherwood, and B.~Calder, ``Using
  simpoint for accurate and efficient simulation,'' in \emph{Proceedings of the
  2003 ACM SIGMETRICS International Conference on Measurement and Modeling of
  Computer Systems}, ser. SIGMETRICS '03, 2003, pp. 318--319.

\bibitem{Zhang:2016:TAS:3001136.3001186}
Y.~Zhang, D.~Meisner, J.~Mars, and L.~Tang, ``Treadmill: Attributing the source
  of tail latency through precise load testing and statistical inference,'' in
  \emph{Proceedings of the 43rd International Symposium on Computer
  Architecture}, ser. ISCA '16, 2016, pp. 456--468.

\bibitem{Lee:2007:MIL:1229428.1229479}
B.~C. Lee, D.~M. Brooks, B.~R. de~Supinski, M.~Schulz, K.~Singh, and S.~A.
  McKee, ``Methods of inference and learning for performance modeling of
  parallel applications,'' in \emph{Proceedings of the 12th ACM SIGPLAN
  Symposium on Principles and Practice of Parallel Programming}, ser. PPoPP
  '07, 2007, pp. 249--258.

\bibitem{Curtis-Maury:2006:OPA:1183401.1183426}
M.~Curtis-Maury, J.~Dzierwa, C.~D. Antonopoulos, and D.~S. Nikolopoulos,
  ``Online power-performance adaptation of multithreaded programs using
  hardware event-based prediction,'' in \emph{Proceedings of the 20th Annual
  International Conference on Supercomputing}, ser. ICS '06, 2006, pp.
  157--166.

\bibitem{6957246}
P.~E. Bailey, D.~K. Lowenthal, V.~Ravi, B.~Rountree, M.~Schulz, and B.~R.
  d.~Supinski, ``Adaptive configuration selection for power-constrained
  heterogeneous systems,'' in \emph{2014 43rd International Conference on
  Parallel Processing}, 2014, pp. 371--380.

\bibitem{Berral:2010:TES:1791314.1791349}
J.~L. Berral, I.~n. Goiri, R.~Nou, F.~Juli\`{a}, J.~Guitart, R.~Gavald\`{a},
  and J.~Torres, ``Towards energy-aware scheduling in data centers using
  machine learning,'' in \emph{Proceedings of the 1st International Conference
  on Energy-Efficient Computing and Networking}, ser. e-Energy '10, 2010, pp.
  215--224.

\bibitem{DelVento:2012:POS:2185475.2185477}
D.~Del~Vento, ``Performance optimization on a supercomputer with ctuning and
  the {PGI} compiler,'' in \emph{Proceedings of the 2Nd International Workshop
  on Adaptive Self-Tuning Computing Systems for the Exaflop Era}, ser. EXADAPT
  '12, 2012, pp. 12--20.

\bibitem{micolet2016machine}
P.-J. Micolet, A.~Smith, and C.~Dubach, ``A machine learning approach to
  mapping streaming workloads to dynamic multicore processors,'' in \emph{ACM
  SIGPLAN Notices}, vol.~51, no.~5, 2016, pp. 113--122.

\bibitem{6494993}
D.~Grewe, Z.~Wang, and M.~F.~P. O'Boyle, ``Portable mapping of data parallel
  programs to {OpenCL} for heterogeneous systems,'' in \emph{Proceedings of the
  2013 IEEE/ACM International Symposium on Code Generation and Optimization
  (CGO)}, 2013, pp. 1--10.

\bibitem{Yu:2000:ARP:335231.335238}
H.~Yu and L.~Rauchwerger, ``Adaptive reduction parallelization techniques,'' in
  \emph{Proceedings of the 14th International Conference on Supercomputing},
  ser. ICS '00, 2000, pp. 66--77.

\bibitem{Leather:2009:AFG:1545006.1545059}
H.~Leather, E.~Bonilla, and M.~O'Boyle, ``Automatic feature generation for
  machine learning based optimizing compilation,'' in \emph{Proceedings of the
  7th Annual IEEE/ACM International Symposium on Code Generation and
  Optimization}, ser. CGO '09, 2009, pp. 81--91.

\bibitem{Wang:2014:APM:2695583.2677036}
Z.~Wang, D.~Grewe, and M.~F.~P. O'boyle, ``Automatic and portable mapping of
  data parallel programs to opencl for {GPU-Based} heterogeneous systems,''
  \emph{ACM Trans. Archit. Code Optim.}, vol.~11, no.~4, pp. 42:1--42:26, 2014.

\bibitem{Ding:2015:AAC:2737924.2737969}
Y.~Ding, J.~Ansel, K.~Veeramachaneni, X.~Shen, U.-M. O'Reilly, and
  S.~Amarasinghe, ``Autotuning algorithmic choice for input sensitivity,'' in
  \emph{Proceedings of the 36th ACM SIGPLAN Conference on Programming Language
  Design and Implementation}, ser. PLDI '15, 2015, pp. 379--390.

\bibitem{Ho:1995:RDF:844379.844681}
T.~K. Ho, ``Random decision forests,'' in \emph{Proceedings of the Third
  International Conference on Document Analysis and Recognition (Volume 1) -
  Volume 1}, ser. ICDAR '95, 1995.

\bibitem{Dietterich:2000:EMM:648054.743935}
T.~G. Dietterich, ``Ensemble methods in machine learning,'' in
  \emph{Proceedings of the First International Workshop on Multiple Classifier
  Systems}, ser. MCS '00, 2000, pp. 1--15.

\bibitem{lokuciejewski2009automatic}
P.~Lokuciejewski, F.~Gedikli, P.~Marwedel, and K.~Morik, ``Automatic {WCET}
  reduction by machine learning based heuristics for function inlining,'' in
  \emph{3rd Workshop on Statistical and Machine Learning Approaches to
  Architectures and Compilation (SMART)}, 2009, pp. 1--15.

\bibitem{7284455}
S.~Benedict, R.~S. Rejitha, P.~Gschwandtner, R.~Prodan, and T.~Fahringer,
  ``Energy prediction of openmp applications using random forest modeling
  approach,'' in \emph{2015 IEEE International Parallel and Distributed
  Processing Symposium Workshop}, 2015, pp. 1251--1260.

\bibitem{rejitha2017energy}
R.~Rejitha, S.~Benedict, S.~A. Alex, and S.~Infanto, ``Energy prediction of
  cuda application instances using dynamic regression models,''
  \emph{Computing}, pp. 1--26, 2017.

\bibitem{pact17}
C.~Cummins, P.~Petoumenos, Z.~Wang, and H.~Leather, ``End-to-end deep learning
  of optimization heuristics,'' in \emph{The 26th International Conference on
  Parallel Architectures and Compilation Techniques (PACT)}, ser. PACT '17,
  2017.

\bibitem{wang2014integrating}
Z.~Wang, G.~Tournavitis, B.~Franke, and M.~F. O'boyle, ``Integrating
  profile-driven parallelism detection and machine-learning-based mapping,''
  \emph{ACM Transactions on Architecture and Code Optimization (TACO)},
  vol.~11, no.~1, p.~2, 2014.

\bibitem{Taylor2017AdaptiveOF}
B.~Taylor, V.~S. Marco, and Z.~Wang, ``Adaptive optimization for {OpenCL}
  programs on embedded heterogeneous systems,'' in \emph{The 18th Annual ACM
  SIGPLAN / SIGBED Conference on Languages, Compilers, and Tools for Embedded
  Systems}, ser. LCETS '17, 2017.

\bibitem{zheng18}
P.~Zhang, J.~Fang, T.~Tang, C.~Yang, and Z.~Wang, ``Auto-tuning streamed
  applications on {Intel Xeon Phi},'' in \emph{32nd IEEE International Parallel
  \& Distributed Processing Symposium}, ser. IPDPS, 2018.

\bibitem{spmv}
S.~Chen, J.~Fang, D.~Chen, C.~Xu, and Z.~Wang, ``Adaptive optimization of
  sparse matrix-vector multiplication on emerging many-core architectures,'' in
  \emph{The 20th IEEE International Conference on High Performance Computing
  and Communications (HPCC)}, 2018.

\bibitem{Joseph:2006:PPM:1194816.1194863}
P.~J. Joseph, K.~Vaswani, and M.~J. Thazhuthaveetil, ``A predictive performance
  model for superscalar processors,'' in \emph{Proceedings of the 39th Annual
  IEEE/ACM International Symposium on Microarchitecture}, ser. MICRO 39, 2006,
  pp. 161--170.

\bibitem{Ganapathi:2009:CML:1855591.1855592}
A.~Ganapathi, K.~Datta, A.~Fox, and D.~Patterson, ``A case for machine learning
  to optimize multicore performance,'' in \emph{Proceedings of the First USENIX
  Conference on Hot Topics in Parallelism}, ser. HotPar'09, 2009.

\bibitem{Deniz2016}
E.~Deniz and A.~Sen, ``Using machine learning techniques to detect parallel
  patterns of multi-threaded applications,'' \emph{International Journal of
  Parallel Programming}, vol.~44, no.~4, pp. 867--900, 2016.

\bibitem{dl}
Y.~LeCun, Y.~Bengio, and G.~Hinton, \emph{Deep Learning}, 2015.

\bibitem{Krizhevsky_imagenetclassification}
A.~Krizhevsky, I.~Sutskever, and G.~E. Hinton, ``Imagenet classification with
  deep convolutional neural networks,'' in \emph{Advances in Neural Information
  Processing Systems (NIPS)}, 2012.

\bibitem{He_2016_CVPR}
K.~He, X.~Zhang, S.~Ren, and J.~Sun, ``Deep residual learning for image
  recognition,'' in \emph{The IEEE Conference on Computer Vision and Pattern
  Recognition (CVPR)}, June 2016.

\bibitem{Lee:2009:UFL:2984093.2984217}
H.~Lee, Y.~Largman, P.~Pham, and A.~Y. Ng, ``Unsupervised feature learning for
  audio classification using convolutional deep belief networks,'' in
  \emph{Proceedings of the 22Nd International Conference on Neural Information
  Processing Systems}, ser. NIPS, 2009, pp. 1096--1104.

\bibitem{Allamanis2017}
M.~Allamanis and C.~Sutton, ``{A Survey of Machine Learning for Big Code and
  Naturalness},'' 2017.

\bibitem{macqueen1967some}
J.~MacQueen \emph{et~al.}, ``Some methods for classification and analysis of
  multivariate observations,'' 1967.

\bibitem{Sherwood:2002:ACL:605397.605403}
T.~Sherwood, E.~Perelman, G.~Hamerly, and B.~Calder, ``Automatically
  characterizing large scale program behavior,'' in \emph{Proceedings of the
  10th International Conference on Architectural Support for Programming
  Languages and Operating Systems}, ser. ASPLOS X, 2002, pp. 45--57.

\bibitem{Wang:2010:PSP:1854273.1854313}
Z.~Wang and M.~F. O'Boyle, ``Partitioning streaming parallelism for
  multi-cores: A machine learning based approach,'' in \emph{Proceedings of the
  19th International Conference on Parallel Architectures and Compilation
  Techniques}, ser. PACT '10, 2010, pp. 307--318.

\bibitem{Newman:2010:NI:1809753}
M.~Newman, \emph{Networks: An Introduction}.\hskip 1em plus 0.5em minus
  0.4em\relax New York, NY, USA: Oxford University Press, Inc., 2010.

\bibitem{Martins:2014:ECO:2597809.2597821}
L.~G. Martins, R.~Nobre, A.~C. Delbem, E.~Marques, and J.~a.~M. Cardoso,
  ``Exploration of compiler optimization sequences using clustering-based
  selection,'' in \emph{Proceedings of the 2014 SIGPLAN/SIGBED Conference on
  Languages, Compilers and Tools for Embedded Systems}, ser. LCTES '14, 2014,
  pp. 63--72.

\bibitem{1106006}
L.~Eeckhout, H.~Vandierendonck, and K.~D. Bosschere, ``Workload design:
  selecting representative program-input pairs,'' in
  \emph{Proceedings.International Conference on Parallel Architectures and
  Compilation Techniques}, 2002, pp. 83--94.

\bibitem{Chen:2010:EIO:1806596.1806647}
Y.~Chen, Y.~Huang, L.~Eeckhout, G.~Fursin, L.~Peng, O.~Temam, and C.~Wu,
  ``Evaluating iterative optimization across 1000 datasets,'' in
  \emph{Proceedings of the 31st ACM SIGPLAN Conference on Programming Language
  Design and Implementation}, ser. PLDI '10, 2010, pp. 448--459.

\bibitem{ashouri2014bayesian}
A.~H. Ashouri, G.~Mariani, G.~Palermo, and C.~Silvano, ``A bayesian network
  approach for compiler auto-tuning for embedded processors,'' in
  \emph{Embedded Systems for Real-time Multimedia (ESTIMedia), 2014 IEEE 12th
  Symposium on}.\hskip 1em plus 0.5em minus 0.4em\relax IEEE, 2014, pp. 90--97.

\bibitem{Phansalkar:2007:ARA:1250662.1250713}
A.~Phansalkar, A.~Joshi, and L.~K. John, ``Analysis of redundancy and
  application balance in the spec cpu2006 benchmark suite,'' in
  \emph{Proceedings of the 34th Annual International Symposium on Computer
  Architecture}, ser. ISCA '07, 2007, pp. 412--423.

\bibitem{Vincent:2008:ECR:1390156.1390294}
P.~Vincent, H.~Larochelle, Y.~Bengio, and P.-A. Manzagol, ``Extracting and
  composing robust features with denoising autoencoders,'' in \emph{Proceedings
  of the 25th International Conference on Machine Learning}, ser. ICML '08,
  2008, pp. 1096--1103.

\bibitem{Gudeepapi}
X.~Gu, H.~Zhang, D.~Zhang, and S.~Kim, ``{Deep API Learning}.''

\bibitem{singer2002learning}
B.~Singer and M.~Veloso, ``Learning to construct fast signal processing
  implementations,'' \emph{Journal of Machine Learning Research}, vol.~3, pp.
  887--919, 2002.

\bibitem{Li:2005:OSG:1048922.1048979}
X.~Li, M.~J. Garzaran, and D.~Padua, ``Optimizing sorting with genetic
  algorithms,'' in \emph{Proceedings of the International Symposium on Code
  Generation and Optimization}, ser. CGO '05, 2005, pp. 99--110.

\bibitem{ansel:pldi:2009}
J.~Ansel, C.~Chan, Y.~L. Wong, M.~Olszewski, Q.~Zhao, A.~Edelman, and
  S.~Amarasinghe, ``Petabricks: A language and compiler for algorithmic
  choice,'' in \emph{ACM SIGPLAN Conference on Programming Language Design and
  Implementation}, ser. PLDI '09, 2009.

\bibitem{Harman:2012:GCC:2351676.2351678}
M.~Harman, W.~B. Langdon, Y.~Jia, D.~R. White, A.~Arcuri, and J.~A. Clark,
  ``The gismoe challenge: Constructing the pareto program surface using genetic
  programming to find better programs (keynote paper),'' in \emph{Proceedings
  of the 27th IEEE/ACM International Conference on Automated Software
  Engineering}, ser. ASE 2012, 2012, pp. 1--14.

\bibitem{Garciarena:2016:EOC:2908961.2931696}
U.~Garciarena and R.~Santana, ``Evolutionary optimization of compiler flag
  selection by learning and exploiting flags interactions,'' in
  \emph{Proceedings of the 2016 on Genetic and Evolutionary Computation
  Conference Companion}, ser. GECCO '16 Companion, 2016, pp. 1159--1166.

\bibitem{6176647}
M.~Zuluaga, E.~Bonilla, and N.~Topham, ``Predicting best design trade-offs: A
  case study in processor customization,'' in \emph{2012 Design, Automation
  Test in Europe Conference Exhibition (DATE)}, 2012, pp. 1030--1035.

\bibitem{jantz2013exploiting}
M.~R. Jantz and P.~A. Kulkarni, ``Exploiting phase inter-dependencies for
  faster iterative compiler optimization phase order searches,'' in
  \emph{Compilers, Architecture and Synthesis for Embedded Systems (CASES),
  2013 International Conference on}.\hskip 1em plus 0.5em minus 0.4em\relax
  IEEE, 2013, pp. 1--10.

\bibitem{ipek2008self}
E.~Ipek, O.~Mutlu, J.~F. Mart{\'\i}nez, and R.~Caruana, ``Self-optimizing
  memory controllers: A reinforcement learning approach,'' in \emph{Computer
  Architecture, 2008. ISCA'08. 35th International Symposium on}.\hskip 1em plus
  0.5em minus 0.4em\relax IEEE, 2008, pp. 39--50.

\bibitem{porter2016rex}
B.~Porter, M.~Grieves, R.~Rodrigues~Filho, and D.~Leslie, ``Rex: A development
  platform and online learning approach for runtime emergent software
  systems,'' in \emph{Symposium on Operating Systems Design and
  Implementation}.\hskip 1em plus 0.5em minus 0.4em\relax USENIX, November
  2016, pp. 333--348.

\bibitem{Rao:2009:VRL:1555228.1555263}
J.~Rao, X.~Bu, C.-Z. Xu, L.~Wang, and G.~Yin, ``Vconf: A reinforcement learning
  approach to virtual machines auto-configuration,'' in \emph{Proceedings of
  the 6th International Conference on Autonomic Computing}, ser. ICAC '09,
  2009, pp. 137--146.

\bibitem{Lagoudakis:2000:ASU:645529.657981}
M.~G. Lagoudakis and M.~L. Littman, ``Algorithm selection using reinforcement
  learning,'' in \emph{Proceedings of the Seventeenth International Conference
  on Machine Learning}, ser. ICML '00, 2000, pp. 511--518.

\bibitem{mishra2018caloree}
N.~Mishra and C.~Imes, ``{CALOREE}: Learning control for predictable latency
  and low energy,'' in \emph{Proceedings of the 23th International Conference
  on Architectural Support for Programming Languages and Operating Systems},
  ser. ASPLOS, 2018.

\bibitem{Emani2014}
M.~K. Emani and M.~O'Boyle, ``Change detection based parallelism mapping:
  Exploiting offline models and online adaptation,'' in \emph{Languages and
  Compilers for Parallel Computing: 27th International Workshop (LCPC 2014)},
  2014, pp. 208--223.

\bibitem{DBLP:journals/corr/Li17b}
Y.~Li, ``Deep reinforcement learning: An overview,'' \emph{CoRR}, vol.
  abs/1701.07274, 2017.

\bibitem{ansel2014opentuner}
J.~Ansel \emph{et~al.}, ``Opentuner: An extensible framework for program
  autotuning,'' in \emph{PACT '14}.

\bibitem{williams1996gaussian}
C.~K. Williams and C.~E. Rasmussen, ``Gaussian processes for regression,'' in
  \emph{Advances in neural information processing systems}, 1996, pp. 514--520.

\bibitem{rasmussen2006gaussian}
C.~E. Rasmussen and C.~K. Williams, \emph{Gaussian processes for machine
  learning}.\hskip 1em plus 0.5em minus 0.4em\relax MIT press Cambridge, 2006,
  vol.~1.

\bibitem{Emani:2015:CDM:2737924.2737999}
M.~K. Emani and M.~O'Boyle, ``Celebrating diversity: A mixture of experts
  approach for runtime mapping in dynamic environments,'' in \emph{Proceedings
  of the 36th ACM SIGPLAN Conference on Programming Language Design and
  Implementation}, ser. PLDI '15, 2015, pp. 499--508.

\bibitem{nguyen2017introduction}
H.~D. Nguyen and F.~Chamroukhi, ``An introduction to the practical and
  theoretical aspects of mixture-of-experts modeling,'' \emph{arXiv preprint
  arXiv:1707.03538}, 2017.

\bibitem{polikar2006ensemble}
R.~Polikar, ``Ensemble based systems in decision making,'' \emph{IEEE Circuits
  and systems magazine}, vol.~6, no.~3, pp. 21--45.

\bibitem{rokach2010ensemble}
L.~Rokach, ``Ensemble-based classifiers,'' \emph{Artificial Intelligence
  Review}, vol.~33, no.~1, pp. 1--39, 2010.

\bibitem{Jiang:2010:ESC:1772954.1772989}
Y.~Jiang, E.~Z. Zhang, K.~Tian, F.~Mao, M.~Gethers, X.~Shen, and Y.~Gao,
  ``Exploiting statistical correlations for proactive prediction of program
  behaviors,'' in \emph{Proceedings of the 8th Annual IEEE/ACM International
  Symposium on Code Generation and Optimization}, ser. CGO '10, 2010, pp.
  248--256.

\bibitem{middleware17}
V.~S. Marco, B.~Taylor, B.~Porter, and Z.~Wang, ``Improving spark application
  throughput via memory aware task co-location: A mixture of experts
  approach,'' in \emph{ACM/IFIP/USENIX Middleware conference}, 2017.

\bibitem{Singer:2000:LPP:645529.657967}
B.~Singer and M.~M. Veloso, ``Learning to predict performance from formula
  modeling and training data,'' in \emph{Proceedings of the Seventeenth
  International Conference on Machine Learning}, ser. ICML '00, 2000, pp.
  887--894.

\bibitem{Park:2012:UGP:2259016.2259042}
E.~Park, J.~Cavazos, and M.~A. Alvarez, ``Using graph-based program
  characterization for predictive modeling,'' in \emph{Proceedings of the Tenth
  International Symposium on Code Generation and Optimization}, ser. CGO '12,
  2012, pp. 196--206.

\bibitem{5708970}
A.~M. Malik, ``Spatial based feature generation for machine learning based
  optimization compilation,'' in \emph{2010 Ninth International Conference on
  Machine Learning and Applications}, 2010, pp. 925--930.

\bibitem{burtscher2012quantitative}
M.~Burtscher, R.~Nasre, and K.~Pingali, ``A quantitative study of irregular
  programs on {GPUs},'' in \emph{Workload Characterization (IISWC), 2012 IEEE
  International Symposium on}, 2012, pp. 141--151.

\bibitem{luo2015fast}
Y.~Luo, G.~Tan, Z.~Mo, and N.~Sun, ``Fast: A fast stencil autotuning framework
  based on an optimal-solution space model,'' in \emph{Proceedings of the 29th
  ACM on International Conference on Supercomputing}, 2015, pp. 187--196.

\bibitem{browne2000portable}
S.~Browne, J.~Dongarra, N.~Garner, G.~Ho, and P.~Mucci, ``A portable
  programming interface for performance evaluation on modern processors,''
  \emph{The international journal of high performance computing applications},
  vol.~14, no.~3, pp. 189--204, 2000.

\bibitem{Mytkowicz:2009:PWD:1508244.1508275}
T.~Mytkowicz, A.~Diwan, M.~Hauswirth, and P.~F. Sweeney, ``Producing wrong data
  without doing anything obviously wrong!'' in \emph{Proceedings of the 14th
  International Conference on Architectural Support for Programming Languages
  and Operating Systems}, ser. ASPLOS XIV, 2009, pp. 265--276.

\bibitem{Cavazos:2006:APM:1176760.1176765}
J.~Cavazos, C.~Dubach, F.~Agakov, E.~Bonilla, M.~F.~P. O'Boyle, G.~Fursin, and
  O.~Temam, ``Automatic performance model construction for the fast software
  exploration of new hardware designs,'' in \emph{Proceedings of the 2006
  International Conference on Compilers, Architecture and Synthesis for
  Embedded Systems}, ser. CASES '06, 2006, pp. 24--34.

\bibitem{khan2007using}
S.~Khan, P.~Xekalakis, J.~Cavazos, and M.~Cintra, ``Using predictivemodeling
  for cross-program design space exploration in multicore systems,'' in
  \emph{Proceedings of the 16th International Conference on Parallel
  Architecture and Compilation Techniques}.\hskip 1em plus 0.5em minus
  0.4em\relax IEEE Computer Society, 2007, pp. 327--338.

\bibitem{Namolaru:2010:PAS:1878921.1878951}
M.~Namolaru, A.~Cohen, G.~Fursin, A.~Zaks, and A.~Freund, ``Practical
  aggregation of semantical program properties for machine learning based
  optimization,'' in \emph{Proceedings of the 2010 International Conference on
  Compilers, Architectures and Synthesis for Embedded Systems}, ser. CASES '10,
  2010, pp. 197--206.

\bibitem{Bishop:2006:PRM:1162264}
C.~M. Bishop, \emph{Pattern Recognition and Machine Learning (Information
  Science and Statistics)}.\hskip 1em plus 0.5em minus 0.4em\relax Secaucus,
  NJ, USA: Springer-Verlag New York, Inc., 2006.

\bibitem{hoste2006performance}
K.~Hoste, A.~Phansalkar, L.~Eeckhout, A.~Georges, L.~K. John, and
  K.~De~Bosschere, ``Performance prediction based on inherent program
  similarity,'' in \emph{Parallel Architectures and Compilation Techniques
  (PACT), 2006 International Conference on}.\hskip 1em plus 0.5em minus
  0.4em\relax IEEE, 2006, pp. 114--122.

\bibitem{Rosenblum:2010:ECP:1806672.1806678}
N.~E. Rosenblum, B.~P. Miller, and X.~Zhu, ``Extracting compiler provenance
  from program binaries,'' in \emph{Proceedings of the 9th ACM SIGPLAN-SIGSOFT
  Workshop on Program Analysis for Software Tools and Engineering}, ser. PASTE
  '10, 2010, pp. 21--28.

\bibitem{7429329}
A.~Bhattacharyya, G.~Kwasniewski, and T.~Hoefler, ``Using compiler techniques
  to improve automatic performance modeling,'' in \emph{2015 International
  Conference on Parallel Architecture and Compilation (PACT)}, 2015, pp.
  468--479.

\bibitem{Taylor:2010:ECH:2898607.2898878}
M.~E. Taylor, K.~E. Coons, B.~Robatmili, B.~A. Maher, D.~Burger, and K.~S.
  McKinley, ``Evolving compiler heuristics to manage communication and
  contention,'' in \emph{Proceedings of the Twenty-Fourth AAAI Conference on
  Artificial Intelligence}, ser. AAAI'10, 2010, pp. 1690--1693.

\bibitem{Ting2016}
P.~Ting, C.~Tu, P.~Chen, Y.~Lo, and S.~Cheng, ``{FEAST: An Automated Feature
  Selection Framework for Compilation Tasks},'' \emph{arXiv:1610.09543}, 2016.

\bibitem{6957226}
E.~Park, C.~Kartsaklis, and J.~Cavazos, ``Hercules: Strong patterns towards
  more intelligent predictive modeling,'' in \emph{43rd International
  Conference on Parallel Processing}, 2014, pp. 172--181.

\bibitem{beyer1999nearest}
K.~Beyer, J.~Goldstein, R.~Ramakrishnan, and U.~Shaft, ``When is ``nearest
  neighbor" meaningful?'' in \emph{International conference on database
  theory}.\hskip 1em plus 0.5em minus 0.4em\relax Springer, 1999, pp. 217--235.

\bibitem{Fodor02asurvey}
I.~Fodor, ``A survey of dimension reduction techniques,'' Lawrence Livermore
  National Laboratory, Tech. Rep., 2002.

\bibitem{thomson2009reducing}
J.~Thomson, M.~F. O'Boyle, G.~Fursin, and B.~Franke, ``Reducing training time
  in a one-shot machine learning-based compiler.'' in \emph{LCPC}, vol.
  5898.\hskip 1em plus 0.5em minus 0.4em\relax Springer, 2009, pp. 399--407.

\bibitem{Bengio:2009:LDA:1658423.1658424}
Y.~Bengio, ``Learning deep architectures for ai,'' \emph{Found. Trends Mach.
  Learn.}, vol.~2, no.~1, pp. 1--127, 2009.

\bibitem{deng2010binary}
L.~Deng, M.~L. Seltzer, D.~Yu, A.~Acero, A.-r. Mohamed, and G.~Hinton, ``Binary
  coding of speech spectrograms using a deep auto-encoder,'' in \emph{Eleventh
  Annual Conference of the International Speech Communication Association},
  2010.

\bibitem{Mou2013}
L.~Mou, G.~Li, L.~Zhang, T.~Wang, and Z.~Jin, ``{Convolutional Neural Networks
  over Tree Structures for Programming Language Processing},'' 2013.

\bibitem{White2016}
M.~White, M.~Tufano, C.~Vendome, and D.~Poshyvanyk, ``{Deep Learning Code
  Fragments for Code Clone Detection},'' in \emph{ASE '16 (31st IEEE/ACM
  International Conference on Automated Software Engineering)}, 2016, pp.
  87--98.

\bibitem{Cummins:2017:SBP:3049832.3049843}
C.~Cummins, P.~Petoumenos, Z.~Wang, and H.~Leather, ``Synthesizing benchmarks
  for predictive modeling,'' in \emph{Proceedings of the 2017 International
  Symposium on Code Generation and Optimization}, ser. CGO '17, 2017, pp.
  86--99.

\bibitem{White2017}
M.~White, M.~Tufano, M.~Mart{\'{i}}nez, M.~Monperrus, and D.~Poshyvanyk,
  ``{Sorting and Transforming Program Repair Ingredients via Deep Learning Code
  Similarities},'' 2017.

\bibitem{Almagor:2004:FEC:997163.997196}
L.~Almagor, K.~D. Cooper, A.~Grosul, T.~J. Harvey, S.~W. Reeves,
  D.~Subramanian, L.~Torczon, and T.~Waterman, ``Finding effective compilation
  sequences,'' in \emph{Proceedings of the 2004 ACM SIGPLAN/SIGBED Conference
  on Languages, Compilers, and Tools for Embedded Systems}, ser. LCTES '04,
  2004, pp. 231--239.

\bibitem{Cooper:2005:AAC:1065910.1065921}
K.~D. Cooper, A.~Grosul, T.~J. Harvey, S.~Reeves, D.~Subramanian, L.~Torczon,
  and T.~Waterman, ``{ACME}: Adaptive compilation made efficient,'' in
  \emph{Proceedings of the 2005 ACM SIGPLAN/SIGBED Conference on Languages,
  Compilers, and Tools for Embedded Systems}, ser. LCTES '05, 2005, pp. 69--77.

\bibitem{Ashouri:2017:MMC:3132652.3124452}
A.~H. Ashouri, A.~Bignoli, G.~Palermo, C.~Silvano, S.~Kulkarni, and J.~Cavazos,
  ``{MiCOMP}: Mitigating the compiler phase-ordering problem using optimization
  sub-sequences and machine learning,'' \emph{ACM Trans. Archit. Code Optim.},
  vol.~14, no.~3, pp. 29:1--29:28, 2017.

\bibitem{cooper2002adaptive}
K.~D. Cooper, D.~Subramanian, and L.~Torczon, ``Adaptive optimizing compilers
  for the 21st century,'' \emph{The Journal of Supercomputing}, vol.~23, no.~1,
  pp. 7--22, 2002.

\bibitem{5688317}
D.~R. White, A.~Arcuri, and J.~A. Clark, ``Evolutionary improvement of
  programs,'' \emph{IEEE Transactions on Evolutionary Computation}, vol.~15,
  no.~4, pp. 515--538, 2011.

\bibitem{Fursin:2010:COP:1880043.1880047}
G.~Fursin and O.~Temam, ``Collective optimization: A practical collaborative
  approach,'' \emph{ACM Trans. Archit. Code Optim.}, vol.~7, no.~4, pp.
  20:1--20:29, Dec. 2010.

\bibitem{Kukunas:2010:GAI:1830761.1830879}
J.~Kukunas, R.~D. Cupper, and G.~M. Kapfhammer, ``A genetic algorithm to
  improve linux kernel performance on resource-constrained devices,'' in
  \emph{Proceedings of the 12th Annual Conference Companion on Genetic and
  Evolutionary Computation}, ser. GECCO '10, 2010, pp. 2095--2096.

\bibitem{Pouchet:2008:IOP:1375581.1375594}
L.-N. Pouchet, C.~Bastoul, A.~Cohen, and J.~Cavazos, ``Iterative optimization
  in the polyhedral model: Part ii, multidimensional time,'' in
  \emph{Proceedings of the 29th ACM SIGPLAN Conference on Programming Language
  Design and Implementation}, ser. PLDI '08, 2008, pp. 90--100.

\bibitem{asanovic2006landscape}
K.~Asanovic, R.~Bodik, B.~C. Catanzaro, J.~J. Gebis, P.~Husbands, K.~Keutzer,
  D.~A. Patterson, W.~L. Plishker, J.~Shalf, S.~W. Williams \emph{et~al.},
  ``The landscape of parallel computing research: A view from berkeley,''
  Technical Report UCB/EECS-2006-183, University of California, Berkeley, Tech.
  Rep., 2006.

\bibitem{1419864}
Y.~Zhang, M.~Voss, and E.~S. Rogers, ``Runtime empirical selection of loop
  schedulers on hyperthreaded smps,'' in \emph{19th IEEE International Parallel
  and Distributed Processing Symposium}, ser. IPDPS '05, 2005.

\bibitem{6298188}
D.~Rughetti, P.~D. Sanzo, B.~Ciciani, and F.~Quaglia, ``Machine learning-based
  self-adjusting concurrency in software transactional memory systems,'' in
  \emph{2012 IEEE 20th International Symposium on Modeling, Analysis and
  Simulation of Computer and Telecommunication Systems}, 2012, pp. 278--285.

\bibitem{Delimitrou:2014:QRQ:2541940.2541941}
C.~Delimitrou and C.~Kozyrakis, ``Quasar: Resource-efficient and qos-aware
  cluster management,'' in \emph{Proceedings of the 19th International
  Conference on Architectural Support for Programming Languages and Operating
  Systems}, ser. ASPLOS '14, 2014, pp. 127--144.

\bibitem{5395311}
X.~Chen and S.~Long, ``Adaptive multi-versioning for openmp parallelization via
  machine learning,'' in \emph{2009 15th International Conference on Parallel
  and Distributed Systems}, ser. ICPADS '09, 2009, pp. 907--912.

\bibitem{6152736}
M.~Castro, L.~F.~W. Góes, C.~P. Ribeiro, M.~Cole, M.~Cintra, and J.~F.
  Méhaut, ``A machine learning-based approach for thread mapping on
  transactional memory applications,'' in \emph{2011 18th International
  Conference on High Performance Computing}, 2011, pp. 1--10.

\bibitem{Jung:2011:BES:1993498.1993509}
C.~Jung, S.~Rus, B.~P. Railing, N.~Clark, and S.~Pande, ``Brainy: Effective
  selection of data structures,'' in \emph{Proceedings of the 32Nd ACM SIGPLAN
  Conference on Programming Language Design and Implementation}, ser. PLDI '11,
  2011, pp. 86--97.

\bibitem{Wang:2013:UML:2509420.2512436}
Z.~Wang and M.~F.~P. O'boyle, ``Using machine learning to partition streaming
  programs,'' \emph{ACM Trans. Archit. Code Optim.}, vol.~10, no.~3, pp.
  20:1--20:25, 2013.

\bibitem{chan:sc:2009}
C.~Chan, J.~Ansel, Y.~L. Wong, S.~Amarasinghe, and A.~Edelman, ``Autotuning
  multigrid with petabricks,'' in \emph{ACM/IEEE Conference on Supercomputing},
  ser. SC '09, 2009.

\bibitem{pacula:evoapps:2012}
M.~Pacula, J.~Ansel, S.~Amarasinghe, and U.-M. O'Reilly, ``Hyperparameter
  tuning in bandit-based adaptive operator selection,'' in \emph{European
  Conference on the Applications of Evolutionary Computation}, ser. EuroSys
  '12, 2012.

\bibitem{Ansel:2012:SOA:2380403.2380425}
J.~Ansel, M.~Pacula, Y.~L. Wong, C.~Chan, M.~Olszewski, U.-M. O'Reilly, and
  S.~Amarasinghe, ``Siblingrivalry: Online autotuning through local
  competitions,'' in \emph{Proceedings of the 2012 International Conference on
  Compilers, Architectures and Synthesis for Embedded Systems}, ser. CASES '12,
  2012, pp. 91--100.

\bibitem{Voss:2001:HAP:379539.379583}
M.~J. Voss and R.~Eigemann, ``High-level adaptive program optimization with
  adapt,'' in \emph{Proceedings of the Eighth ACM SIGPLAN Symposium on
  Principles and Practices of Parallel Programming}, ser. PPoPP '01, 2001, pp.
  93--102.

\bibitem{tiwari2011online}
A.~Tiwari and J.~K. Hollingsworth, ``Online adaptive code generation and
  tuning,'' in \emph{IEEE International Parallel \& Distributed Processing
  Symposium (IPDPS)}, 2011, pp. 879--892.

\bibitem{infocom17}
J.~Ren, L.~Gao, H.~Wang, and Z.~Wang, ``Optimise web browsing on heterogeneous
  mobile platforms: a machine learning based approach,'' in \emph{IEEE
  International Conference on Computer Communications (INFOCOM), 2017}, ser.
  INFOCOM 2017, 2017.

\bibitem{zhu2013high}
Y.~Zhu and V.~J. Reddi, ``High-performance and energy-efficient mobile web
  browsing on big/little systems,'' ser. HPCA '13, 2013.

\bibitem{Wang:2013:SAM:2495258.2495918}
Z.~Wang, M.~F.~P. O'Boyle, and M.~K. Emani, ``Smart, adaptive mapping of
  parallelism in the presence of external workload,'' in \emph{Proceedings of
  the 2013 IEEE/ACM International Symposium on Code Generation and Optimization
  (CGO)}, ser. CGO '13, 2013, pp. 1--10.

\bibitem{Grewe:2011:WMA:1944862.1944881}
D.~Grewe, Z.~Wang, and M.~F.~P. O'Boyle, ``A workload-aware mapping approach
  for data-parallel programs,'' in \emph{Proceedings of the 6th International
  Conference on High Performance and Embedded Architectures and Compilers},
  ser. HiPEAC '11, 2011, pp. 117--126.

\bibitem{grewe2013opencl}
D.~Grewe, Z.~Wang, and M.~F. O’Boyle, ``{OpenCL} task partitioning in the
  presence of {GPU} contention,'' in \emph{International Workshop on Languages
  and Compilers for Parallel Computing}.\hskip 1em plus 0.5em minus 0.4em\relax
  Springer, 2013, pp. 87--101.

\bibitem{Tang:2012:CNM:2259016.2259018}
L.~Tang, J.~Mars, and M.~L. Soffa, ``Compiling for niceness: Mitigating
  contention for qos in warehouse scale computers,'' in \emph{Proceedings of
  the Tenth International Symposium on Code Generation and Optimization}, ser.
  CGO '12, 2012, pp. 1--12.

\bibitem{Tang:2013:RRS:2451116.2451126}
L.~Tang, J.~Mars, W.~Wang, T.~Dey, and M.~L. Soffa, ``Reqos: Reactive
  static/dynamic compilation for qos in warehouse scale computers,'' in
  \emph{Proceedings of the Eighteenth International Conference on Architectural
  Support for Programming Languages and Operating Systems}, ser. ASPLOS '13,
  2013, pp. 89--100.

\bibitem{Matsunaga:2010:UML:1844765.1845166}
A.~Matsunaga and J.~A.~B. Fortes, ``On the use of machine learning to predict
  the time and resources consumed by applications,'' in \emph{Proceedings of
  the 2010 10th IEEE/ACM International Conference on Cluster, Cloud and Grid
  Computing}, ser. CCGRID '10, 2010, pp. 495--504.

\bibitem{venkataraman2016ernest}
S.~Venkataraman, Z.~Yang, M.~J. Franklin, B.~Recht, and I.~Stoica, ``Ernest:
  Efficient performance prediction for large-scale advanced analytics.'' in
  \emph{NSDI}, 2016, pp. 363--378.

\bibitem{Sankaran:2016:PMB:2903150.2911714}
S.~Sankaran, ``Predictive modeling based power estimation for embedded
  multicore systems,'' in \emph{Proceedings of the ACM International Conference
  on Computing Frontiers}, ser. CF '16, 2016, pp. 370--375.

\bibitem{Zhang:2014:SPQ:2742155.2742197}
Y.~Zhang, M.~A. Laurenzano, J.~Mars, and L.~Tang, ``Smite: Precise qos
  prediction on real-system smt processors to improve utilization in warehouse
  scale computers,'' in \emph{Proceedings of the 47th Annual IEEE/ACM
  International Symposium on Microarchitecture}, ser. MICRO-47, 2014, pp.
  406--418.

\bibitem{petrucci2015octopus}
V.~Petrucci, M.~A. Laurenzano, J.~Doherty, Y.~Zhang, D.~Mosse, J.~Mars, and
  L.~Tang, ``Octopus-man: Qos-driven task management for heterogeneous
  multicores in warehouse-scale computers,'' in \emph{2015 IEEE 21st
  International Symposium on High Performance Computer Architecture
  (HPCA)}.\hskip 1em plus 0.5em minus 0.4em\relax IEEE, 2015, pp. 246--258.

\bibitem{yadwadkar2016multi}
N.~J. Yadwadkar, B.~Hariharan, J.~E. Gonzalez, and R.~Katz, ``Multi-task
  learning for straggler avoiding predictive job scheduling,'' \emph{The
  Journal of Machine Learning Research}, vol.~17, no.~1, pp. 3692--3728, 2016.

\bibitem{David:2014:TCS:2594291.2594343}
Y.~David and E.~Yahav, ``Tracelet-based code search in executables,'' in
  \emph{Proceedings of the 35th ACM SIGPLAN Conference on Programming Language
  Design and Implementation}, ser. PLDI '14, 2014, pp. 349--360.

\bibitem{David:2016:SSB:2908080.2908126}
Y.~David, N.~Partush, and E.~Yahav, ``Statistical similarity of binaries,'' in
  \emph{Proceedings of the 37th ACM SIGPLAN Conference on Programming Language
  Design and Implementation}, ser. PLDI '16, 2016, pp. 266--280.

\bibitem{wong2015clocom}
E.~Wong, T.~Liu, and L.~Tan, ``Clocom: Mining existing source code for
  automatic comment generation,'' in \emph{Software Analysis, Evolution and
  Reengineering (SANER), 2015 IEEE 22nd International Conference on}, 2015, pp.
  380--389.

\bibitem{Fowkes:2016:PPA:2950290.2950319}
J.~Fowkes and C.~Sutton, ``Parameter-free probabilistic api mining across
  github,'' in \emph{Proceedings of the 2016 24th ACM SIGSOFT International
  Symposium on Foundations of Software Engineering}, ser. FSE 2016, 2016, pp.
  254--265.

\bibitem{Nguyen:2016:ACR:2950290.2950333}
A.~T. Nguyen, M.~Hilton, M.~Codoban, H.~A. Nguyen, L.~Mast, E.~Rademacher,
  T.~N. Nguyen, and D.~Dig, ``Api code recommendation using statistical
  learning from fine-grained changes,'' in \emph{Proceedings of the 2016 24th
  ACM SIGSOFT International Symposium on Foundations of Software Engineering},
  ser. FSE 2016, 2016, pp. 511--522.

\bibitem{Raychev:2016:PMC:2983990.2984041}
V.~Raychev, P.~Bielik, and M.~Vechev, ``Probabilistic model for code with
  decision trees,'' in \emph{Proceedings of the 2016 ACM SIGPLAN International
  Conference on Object-Oriented Programming, Systems, Languages, and
  Applications}, ser. OOPSLA 2016, 2016, pp. 731--747.

\bibitem{Bichsel:2016:SDA:2976749.2978422}
B.~Bichsel, V.~Raychev, P.~Tsankov, and M.~Vechev, ``Statistical deobfuscation
  of android applications,'' in \emph{Proceedings of the 2016 ACM SIGSAC
  Conference on Computer and Communications Security}, ser. CCS '16, 2016, pp.
  343--355.

\bibitem{balaprakash2013active}
P.~Balaprakash, R.~B. Gramacy, and S.~M. Wild, ``Active-learning-based
  surrogate models for empirical performance tuning,'' in \emph{Cluster
  Computing (CLUSTER), 2013 IEEE International Conference on}.\hskip 1em plus
  0.5em minus 0.4em\relax IEEE, 2013, pp. 1--8.

\bibitem{ogilvie2014fast}
W.~F. Ogilvie, P.~Petoumenos, Z.~Wang, and H.~Leather, ``Fast automatic
  heuristic construction using active learning,'' in \emph{International
  Workshop on Languages and Compilers for Parallel Computing}, 2014, pp.
  146--160.

\bibitem{zuluaga2013active}
M.~Zuluaga, G.~Sergent, A.~Krause, and M.~P{\"u}schel, ``Active learning for
  multi-objective optimization,'' in \emph{International Conference on Machine
  Learning}, 2013, pp. 462--470.

\bibitem{Ogilvie:2017:MCI:3049832.3049859}
W.~F. Ogilvie, P.~Petoumenos, Z.~Wang, and H.~Leather, ``Minimizing the cost of
  iterative compilation with active learning,'' in \emph{Proceedings of the
  2017 International Symposium on Code Generation and Optimization}, ser. CGO
  '17, 2017, pp. 245--256.

\bibitem{AEC}
\BIBentryALTinterwordspacing
A.~Evaluation. About artifact evaluation. [Online]. Available:
  \url{http://www.artifact-eval.org/about.html}
\BIBentrySTDinterwordspacing

\bibitem{AEC2}
\BIBentryALTinterwordspacing
cTuning Foundation. Artifact evaluation for computer systems' research.
  [Online]. Available: \url{http://ctuning.org/ae/}
\BIBentrySTDinterwordspacing

\end{thebibliography}
